\documentclass[a4paper,11pt]{article}
\pdfoutput=1 

\usepackage{jinstpub} 

\usepackage{siunitx}[=v2]
\DeclareSIUnit\clight{\text{\ensuremath{c}}}
\usepackage[caption=false,font=footnotesize]{subfig}
\usepackage{tikz}
\usetikzlibrary{fadings}
\usetikzlibrary{patterns}
\usetikzlibrary{shadows.blur}
\usetikzlibrary{shapes}
\title{\boldmath Evaluation of Planar Silicon Pixel Sensors with the RD53A Readout Chip for the Phase-2 Upgrade of the CMS Inner Tracker}

\author[1]{The Tracker Group of the CMS Collaboration\note{Corresponding author: 
Georg Steinbrück}}

\emailAdd{georg.steinbrueck@uni-hamburg.de}

\abstract{
    The Large Hadron Collider at CERN will undergo an upgrade in order to increase its luminosity to \SI{7.5e34}{cm^{-2}s^{-1}}. The increased luminosity during this High-Luminosity running phase, starting around 2029, means a higher rate of proton-proton interactions, hence a larger ionizing dose and particle fluence for the detectors.
    The current tracking system of the CMS experiment will be fully replaced in order to cope with the new  operating conditions. Prototype planar pixel sensors for the CMS Inner Tracker with square \SI{50x50}{\micro\meter} and rectangular \SI{100x25}{\micro\meter} pixels read out by the RD53A chip were characterized in the lab and at the DESY-II testbeam facility in order to identify designs that meet the requirements of CMS during the High-Luminosity running phase. A spatial resolution of approximately \SI{3.4}{\micro\meter} (\SI{2}{\micro\meter}) is obtained using the modules with  \SI{50x50}{\micro\meter} (\SI{100x25}{\micro\meter}) pixels at the optimal angle of incidence before irradiation. After irradiation to a \SI{1}{MeV} neutron equivalent fluence of $\Phi_{\rm eq} =  $ \SI{5.3e15}{\per\centi\meter\squared}, a resolution of \SI{9.4}{\micro\meter} is achieved at a bias voltage of \SI{800}{V} using a module with \SI{50x50}{\micro\meter} pixel size. All modules retain a hit efficiency in excess of 99\% after irradiation to fluences up to \SI{2.1e16}{\per\centi\meter\squared}.
Further studies of the electrical properties of the modules, especially crosstalk, are also presented in this paper.}


\begin{document}
\maketitle
\flushbottom





\newpage
\section{Introduction}
The Large Hadron Collider (LHC) at CERN will be upgraded in order to increase the instantaneous luminosity from currently \SI{2e34}{\per\centi\meter\squared\per\second} to \SI{7.5e34}{\per\centi\meter\squared\per\second}, 
boosting the physics potential of its experiments~\cite{hllhc}. An integrated luminosity of \SI{3000} to \SI{4000}{\per\femto\barn} will have been delivered by the end of the 10 year High-Luminosity LHC (HL-LHC) program, which is an increase by about a factor of ten with respect to the first three runs of the LHC ending in 2025~\cite{hllhc2}.
The increased instantaneous luminosity means a higher rate of proton-proton interactions, on the order of 200 per bunch crossing, and thus a higher particle fluence and total ionizing dose (TID) in the detectors. The Compact Muon Solenoid (CMS) detector~\cite{CMS_2008, Adam_2021}  will be upgraded 
in order to maintain or even improve its measurement capabilities under such challenging conditions.

\begin{figure}[!b]
\centering
\includegraphics[width=1\linewidth]{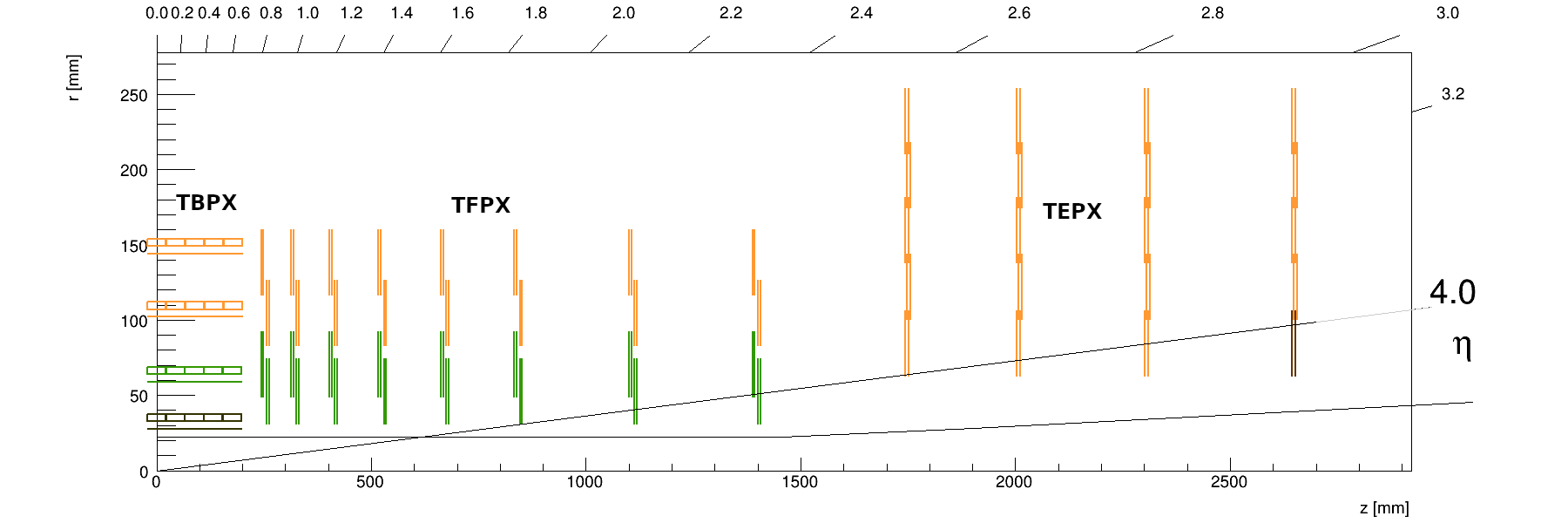}
    \caption{Layout of the CMS Inner Tracker (IT) for Phase-2. In the IT, pixel detector modules with 2$\times$2 readout chips (shown in yellow) and 1$\times$2 readout chips (shown in green) are used~\cite{cms:tkupgrade}. The innermost part of TBPX (shown in black) comprises 1$\times$2 modules with two individual 3D sensors. The innermost ring of disc~4 of TEPX (shown in brown) consists of 2$\times$2 modules and sends data to the luminosity system. The lower black line represents the outer radius of the beam pipe.}
\label{ITLayout}
\end{figure}

This paper focuses on the upgrade of the CMS Inner Tracker (IT)~\cite{cms:tkupgrade}, which is entirely composed of silicon pixel detectors. 
The upgraded IT will feature a two-phase CO$_2$ cooling system with a liquid CO$_2$ temperature of \SI{-33}{\celsius}, leading to sensor temperatures in the range of \SI{-20}{}  to  \SI{-15}{\celsius}.
A layout of the CMS IT is shown in Fig.~\ref{ITLayout}. 
The IT is constructed in three parts: the Tracker Barrel Pixel Detector (TBPX), the Tracker Forward Pixel Detector (TFPX), and the Tracker Endcap Pixel Detector (TEPX). The TBPX comprises four cylindrical detector layers, while the TFPX and TEPX feature eight and four discs per side, respectively. 

%
%
%
The innermost layer of the IT barrel section will be located at a radius of \SI{28}{\milli\meter} from the beam axis. Here, a 
non-ionizing energy loss (NIEL) corresponding to a 
\SI{1}{MeV} neutron equivalent particle fluence of $\Phi_{\rm eq} = $ \SI{3.5e16}{\per\centi\meter\squared} and a total ionizing dose (TID) of \SI{19}{\mega\gray} are expected after \SI{4000}{\per\femto\barn} of integrated luminosity over 10 years of operation. Neither the readout chips nor the sensors are 
expected to be operable under these conditions and a replacement of the first layer is foreseen part-way through the HL-LHC running period.
The radial distances 
of all barrel layers and of the rings of the forward and endcap discs from the center of the beam pipe are reported in Ref.~\cite{cms:tkupgrade}. 

Planar n\textsuperscript{+}-p pixel sensors with an active thickness of \SI{150}{\micro\meter} will be used throughout the IT with the exception of the first layer of TBPX,   
where 3D sensors are the baseline choice given their intrinsically higher radiation tolerance and lower power consumption~\cite{DUARTECAMPDERROS2019162625}. 
Owing to their more complicated production process, which results in a lower production yield and higher prices, 3D sensors are not an option for the entire IT. 
%

The maximum fluence for planar sensors will be reached in ring 1 of the TFPX. For the full lifetime of the IT, with \SI{4000}{\per\femto\barn} delivered, the fluence in this ring is expected to reach \SI{2.3e16}{\per\centi\meter\squared}. 
This value refers to the maximum fluence, received at the inner module edge, while the mean fluence over the module is much lower, about \SI{1.3e16}{\per\centi\meter\squared} over the full detector lifetime.
In ring 2 of TFPX and barrel layer 2, fluences of  \SI{1.1e16}{\per\centi\meter\squared} and \SI{9.4e15}{\per\centi\meter\squared} are expected, respectively.
The IT is constructed such that ring 1 in TFPX can be exchanged at a fluence of  about \SI{1.2e16}{\per\centi\meter\squared} part-way through the HL-LHC period.
In this case, the limiting factor for the lifetime is likely the increase 
in power consumption of the sensors, 
leading to a deterioration of the cooling performance, and ultimately thermal runaway.
The CMS readout chip has been tested up to a total ionizing dose of 
\SI{10}{MGy}. Tests at the dose level of 
\SI{15}{MGy}, expected for the detector region equipped with planar sensors for the full detector lifetime, are planned for 2023.
%
%

Initially, pixel sizes of \SI{50x50}{\micro\meter} and \SI{100x25}{\micro\meter}
were considered, with the long side oriented parallel to the beam axis (along $z$) in the barrel and radially outwards (along $r$) in the forward  and endcap discs.
In the process of evaluating pixel tracking performance, sensors with both pixel sizes were measured.
Guided by tracking and physics simulations of the entire tracking system, CMS has decided to use \SI{100x25}{\micro\meter} pitch sensors throughout the entire IT (including the 3D modules in barrel layer 1).
%

A variety of prototype pixel sensors have been designed and fabricated. These include 3D and planar sensors of both pixel sizes with p-stop or p-spray isolation and with or without a biasing scheme for sensor tests before flip-chip bump bonding~\cite{cms:sensor}. The sensors evaluated in this paper are planar sensors designed to match the layout of the RD53A readout chip (ROC), which is a half-size prototype of the final ROC developed by the RD53 Collaboration, a joint effort by the ATLAS and CMS Collaborations~\cite{rd53a:manual}.
The measurements are performed using the \emph{Linear front-end}~\cite{lfepaper} of the RD53A chip.

This paper describes studies for a variety of RD53A-sensor assemblies 
with the aim to evaluate their performance and to select sensor designs that meet CMS specifications. A list of selected specifications for planar pixel sensors is shown in Table \ref{table_specs}. Since tuning of the chip parameters is critical for an optimal performance of the assemblies and to obtain reliable results, a significant part of the paper is devoted to this aspect.

The pixel modules studied in this paper are described in Section \ref{modules}. In Section \ref{tuning}, the tuning procedure and performance of the RD53A-sensor assemblies are described. 
Laboratory measurements of crosstalk in these modules are reported in Section \ref{xtalk}. Evaluation of the sensors with particle beams and the results are discussed in Section \ref{testbeam}.

\begin{table}[ht]
  \centering
  \caption{Selected requirements for planar pixel sensors used for this measurement and vendor qualification campaign. The requirements for the procurement of sensors for the final experiment differ slightly.
  The full depletion voltage and the hit efficiency  are denoted by V$_{\rm depl}$ and $\epsilon_{\rm{hit}}$,
  \label{table_specs}
  respectively.}
   \begin{tabular}{l l l}
   Parameter &Value & Measured at \\
      \hline
      \hline
   polarity & n$^+$-p &  \\
   \hline
       active thickness & \SI{150}{\micro\meter} &  \\
      \hline
    bulk resistivity & \SIrange{3}{20}{k\ohm\centi\meter} & room temperature \\
      \hline  
   pixel size  & \SI{50x50}{\micro\meter} \\&\SI{100x25}{\micro\meter} \\
    \hline      
       breakdown voltage & $>$\SI{300}{\volt} & \\
        before irradiation & &  \\
      \hline
          breakdown voltage & $>$\SI{800}{\volt} & \\
         after $\Phi_{\rm eq} =  
       $ \SI{5e15}{\per\centi\meter\squared} & &  \\
      \hline
       leakage current & $<$\SI{0.75}{\micro\ampere\per\centi\meter\squared} &   at V$_{\rm depl}$ + \SI{50}{\volt}  \\
     before irradiation & & at \SI{20}{\celsius}  \\
       \hline
      leakage current & $<$\SI{45}{\micro\ampere\per\centi\meter\squared} & at \SI{600}{\volt} \\
     after $\Phi_{\rm eq} =  
       $ \SI{5e15}{\per\centi\meter\squared} &  & at \SI{-25}{\celsius} \\
    annealing at \SI{60}{\celsius} for 1 hr&  &  \\


            \hline
       $\epsilon_{\rm{hit}}$ before irradiation & >99\% & $V_{\rm depl}$ + \SI{50}{\volt} \\
         \hline
       $\epsilon_{\rm{hit}}$ for $\Phi_{\rm eq} =  
       $ \SI{5e15}{\per\centi\meter\squared} \qquad & >99\% & $\le$\SI{800}{\volt}, \SI{-20}{\celsius} \\
        \hline
       $\epsilon_{\rm{hit}}$ for $\Phi_{\rm eq} =$ \SI{1e16}{\per\centi\meter\squared} & >98\% & $\le$\SI{800}{\volt}, \SI{-20}{\celsius} \\
   \hline
  \end{tabular}
\end{table}


\section{Pixel Modules}\label{modules}
\begin{figure}[!b]
  \includegraphics[width=0.1\linewidth]{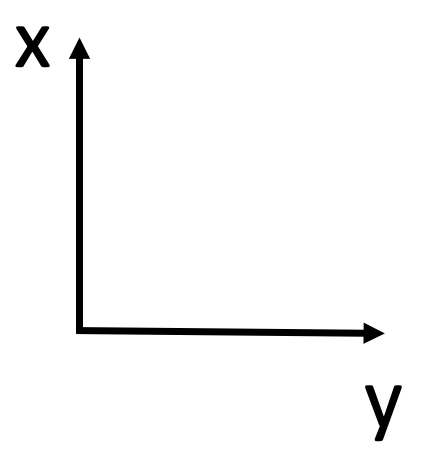}
  \subfloat[]{\includegraphics[width=0.22\linewidth]{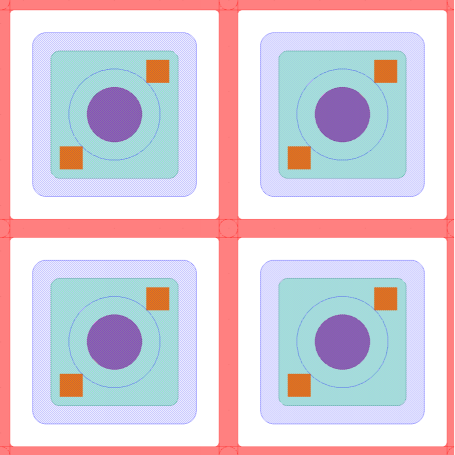}
  \label{fig_rd53a-50p1}}
  \subfloat[]{\includegraphics[width=0.22\linewidth]{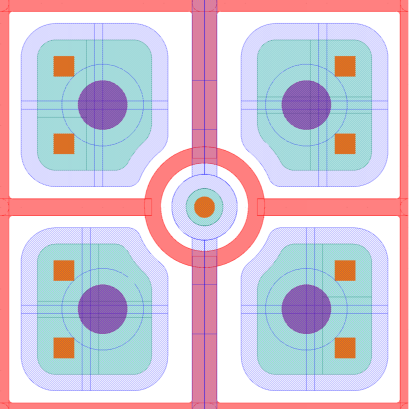}
  \label{fig_rd53a-50p3}}
  \subfloat[]{%
  \begin{minipage}[b][][t]{.44\textwidth}
  \centering
  \includegraphics[width=0.98\linewidth]{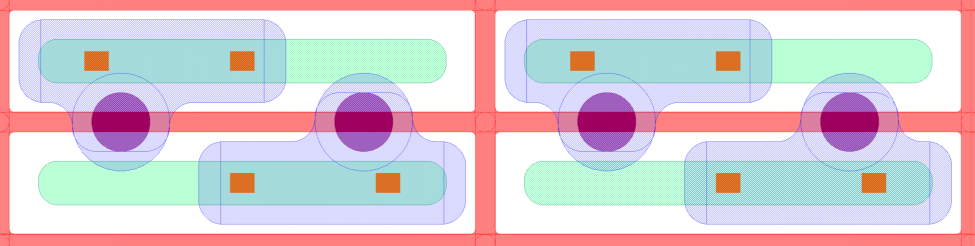}
  \vfill
  \includegraphics[width=0.98\linewidth]{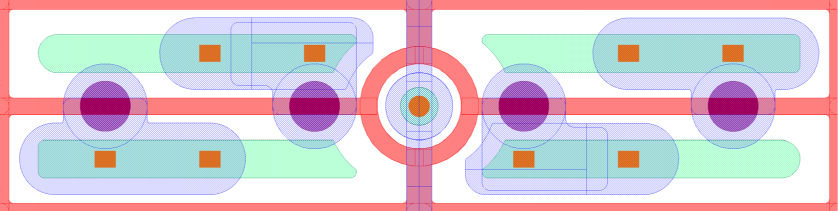}
  \label{fig_rd53a-100}
  \end{minipage}}
    \caption{The layout of a \SI{50x50}{\micro\meter} pitch sensor (a) without and (b) with a bias dot. The corresponding layouts for \SI{100x25}{\micro\meter} pitch sensors are shown in (c). 
    The color code indicates the various mask layers: 
    n$^+$ implant (green), p-stop implant (red), metal contact via (filled orange squares), metallization (blue), and opening in the passivation for bump bonding (purple).}
  \label{fig_layouts}
\end{figure}

A pixel module consists of a silicon pixel sensor bump bonded to one or more readout chips to form a hybrid pixel detector. In this study, only single-chip modules are tested. The baseline choice for the three outer barrel layers and the discs of the IT are fine-pitch n\textsuperscript{+}-p planar pixel sensors with an active thickness of \SI{150}{\micro\meter}. The layouts of the investigated sensor cells are displayed in Fig.~\ref{fig_layouts} together with the definition of the local $x$ and $y$ sensor coordinates used in this paper.

Sensors with and without a punch-through bias dot~\cite{biasdot} are evaluated. The dot enables the biasing of the sensor for testing before bump bonding to a readout chip. 
This makes it possible to identify sensors with low breakdown voltages through current-voltage measurements before the non-reversible bump bonding process.
The prototype sensors discussed in this paper were produced by Hamamatsu Photonics K.K.~\cite{hamamatsu},
one of the three vendors that were qualified by CMS for planar pixel sensors. The details of the sensor design and production are described in Refs.~\cite{cms:sensor, cms:sensor2}.

The sensors were bump bonded to RD53A readout chips with SnAg bumps at the Fraunhofer Institute for Reliability and Microintegration (IZM) in Berlin, Germany~\cite{izm}. The RD53A chip measures \SI{20.0x11.6}{mm} and is produced in \SI{65}{nm} CMOS technology at Taiwan Semiconductor Manufacturing Company, Ltd. (TSMC)~\cite{tsmc}. The \SI{50x50}{\micro\meter} cells of the chip are arranged in a matrix of 400 columns and 192 rows. 
To interconnect a sensor with \SI{100x25}{\micro\meter} pixels to 
the RD53A chip, a metal routing from the implants to the bump bond pads on the sensors is required.

To be able to evaluate various readout schemes at a reduced production cost, the RD53A chip contains three different analog front-end circuits that are designed to meet the HL-LHC requirements in terms of radiation hardness, high hit rate, and stable operation at low thresholds.
The first 128 columns of the ROC are equipped with the \emph{Synchronous front-end}, the next 136 columns with the \emph{Linear front-end}~\cite{linfe}, and the last 136 columns with the \emph{Differential front-end}. All the studies presented in this paper are performed using the \emph{Linear front-end} (LIN FE), as this has been chosen for the final CMS IT readout chip~\cite{lfepaper}.

The IT sensor R\&D program is focused on the evaluation of prototype pixel sensor assemblies with the RD53A chip, with an emphasis on their performance for radiation fluences reached at the end of the HL-LHC running phase. Pixel modules of various sensor designs have been irradiated to fluences in the 
range 
$\Phi_{\rm eq} = $ \SI{4.4e15}{\per\centi\meter\squared} 
to 
\SI{2.1e16}{\per\centi\meter\squared} 
using \SI{25}{\mega\electronvolt} protons at the ZAG Zyklotron AG in Karlsruhe, Germany~\cite{zag} or using \SI{27}{\mega\electronvolt} protons at the MC40 cyclotron facility at the University of Birmingham~\cite{mc40}. The readout chip was not powered during irradiation.
The sensors have not been subjected to any additional annealing other than 
during short periods of handling and transport at room temperature.
A list of the modules studied in this paper is presented in Table \ref{table_modules}.
\begin{table}[!ht]
\centering
\caption{Pixel modules studied in this paper. Modules M563i and M564i were irradiated at the MC40 cyclotron facility in Birmingham while the others were irradiated at the ZAG Zyklotron AG in Karlsruhe. 
The numbers in the last column refer to relevant figure numbers in this paper.}
\bigskip
\label{table_modules}
\begin{tabular}{cccccc}
    ID & \parbox[t]{1cm}{\centering Pitch\newline(\SI{}{\micro\meter\squared})} & Bias dot & \parbox[t]{1.5cm}{\centering $\Phi_{\rm eq} $\newline(\SI{}{\per\centi\meter\squared})} & \parbox[t]{2cm}{Threshold\newline\centering(e)} & Figure\\ \hline\hline
    \multicolumn{6}{c}{Crosstalk measurements} \\ \hline
    M529  & \num{100x25}  & Yes & \num{0} & 900 & \ref{fig_xt} \\ \hline
    \multicolumn{6}{c}{Efficiency measurements} \\ \hline
    M547  & \num{50x50}  & No  & \num{4.4e15} & 1200 & \ref{fig_compEff}\\
    M564i & \num{50x50}  & No  & \num{5.4e15} & 1200 & \ref{fig_compEff}\\
    M542  & \num{100x25} & No  & \num{7.4e15} & 1320 & \ref{fig_compEff}\\
    M563i & \num{100x25} & No  & \num{1.0e16} & 1200 & \ref{fig_compEff}\\
    M589  & \num{100x25} & No  & \num{2.1e16} & 1240 & \ref{fig_compEff}\\ \hline
    \multicolumn{6}{c}{Bias dot measurements} \\ \hline
    M521  & \num{50x50}  & Yes & \num{5.3e15} & 1230 & \ref{fig_m521-eff} \\ \hline
    \multicolumn{6}{c}{Resolution measurements} \\ \hline
    M563  & \num{100x25} & No  & 0 & 700 & \ref{fig_compRes}\\
    M564  & \num{50x50}  & No  & 0 & 600 & \ref{fig_compRes}\\
    M521  & \num{50x50}  & Yes & \num{5.3e15} & 1230 & \ref{fig_compRes}\\ \hline
\end{tabular}
\end{table}

The particle energies correspond to the kinetic energies at extraction from the cyclotrons and a drop of approximately \SI{2}{\mega\electronvolt} is expected during the beam delivery to the devices. 
While the average energy of charged particle primaries in CMS is 
much larger than \SI{23}{\mega\electronvolt}, NIEL scaling~\cite{NIEL} makes it 
possible to compare the effects of bulk radiation damage at different particle 
energies. 
The NIEL scaling hypothesis assumes that the concentration of irradiation induced bulk defects depends only on the NIEL. 
The measured \SI{1}{\mega\electronvolt} neutron equivalent hardness factor for \SI{23}{\mega\electronvolt} protons is $2.2\pm0.4$~\cite{Allport:2019kvs}.
Irradiations of CMS pixel prototype modules with higher energy protons, resulting in a much reduced TID in the readout chip, are under way for comparison.

\section{Tuning and Operation of RD53A Modules}\label{tuning}

The BDAQ53 data acquisition system was used to configure, tune, and operate the RD53A modules~\cite{bdaq53}. A Xilinx KC705 evaluation board~\cite{xilinx} was employed as the hardware platform for the system. The software part of the BDAQ53 is developed in the Python programming language and provides a set of scripts for tuning and measurement purposes.

The RD53A ROC is designed for serial powering,
which is the chosen powering scheme for the CMS IT Phase-2 Upgrade \cite{PEROVIC2020164436}.
However, the RD53A chip also offers the possibility of bypassing the internal power regulator to directly supply power from external units. 
The latter powering scheme was used in the measurements presented in this paper.

\begin{figure}[ht]
\centering
\includegraphics[width=1\linewidth]{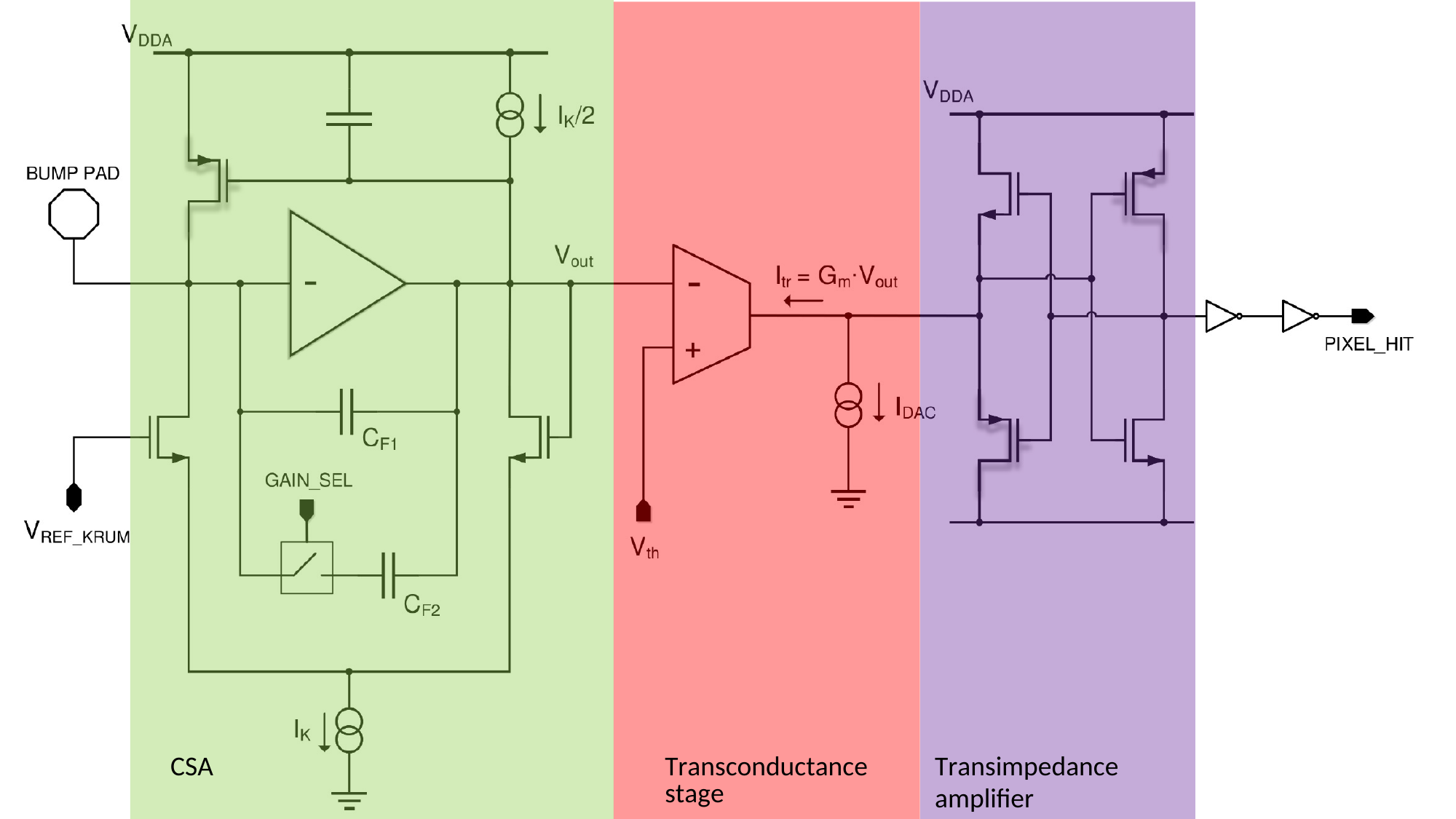}
    \caption{Schematic diagram of the RD53A \emph{Linear front-end}~\cite{rd53a:manual}.}
\label{fig_linfe}
\end{figure}

A schematic diagram of the LIN FE is shown in Fig.~\ref{fig_linfe}. 
The readout chain begins with a low-power charge sensitive amplifier (CSA) featuring a Krummenacher feedback circuit in order to cope with the increased leakage current of the sensor after irradiation. The output of the CSA is digitized using a high-speed current comparator and transimpedance amplifier, and is then processed by the digital pixel logic including a 4-bit Time-over-Threshold (ToT) counter (not shown in the figure).
The global threshold for all pixels is applied to the $V_{\rm th}$ input of the comparator.
A 4-bit binary weighted digital-to-analog converter (DAC) allows for per-pixel threshold adjustment (trimming DAC, TDAC). All pixels of the RD53A chip are equipped with an individual charge injection circuit for test and calibration purposes.

The LIN FE is configured using a number of DAC settings. A short summary of the meaning of the main DAC registers, taken from Ref.~\cite{linfe}, is given here: 
\begin{itemize}
    
     \item Vthreshold\_LIN sets  the  global  threshold  of  the  \emph{Linear front-end},  corresponding  to  the  DC threshold voltage $V_{\rm th}$ applied to the discriminator input.  Increasing Vthreshold\_LIN results in an increased global threshold.
    
     
      \item LDAC\_LIN  sets  the  output dynamic range of the in-pixel threshold trimming DAC (TDAC) that determines the current $I_{\rm DAC}$. Increasing LDAC\_LIN results in an increased output range of the threshold adjustment.

     \item  KRUM\_CURR\_LIN sets the current in the Krummenacher feedback $I_{\rm K}$, used to discharge  the  preamplifier  feedback  capacitance with a constant current.   Increasing  KRUM\_CURR\_LIN  results  in a  faster  preamplifier  return  to  baseline  and  a  reduced ToT.
    \item FC\_BIAS\_LIN sets the current in the preamplifier folded cascode branch.
    
    \item COMP\_LIN sets the bias current in the threshold discriminator input  
    stage.  
    
    \item PA\_IN\_BIAS\_LIN  sets  the  current  in  the  preamplifier  input  branch.   This  current represents  the  main  contribution  to  the \emph{Linear front-end} current  consumption.   A decrease of PA\_IN\_BIAS\_LIN reduces  the chip power,
    at the cost of increased noise.
    
    \item REF\_KRUM\_LIN sets the preamplifier output DC baseline.
\end{itemize}

The values of the DAC parameters used for the operation of the modules are listed in Table~\ref{table_linfe}. The initial values for the DACs are based on recommendations from the front-end designers. In addition to the global threshold mentioned above, 
the current in the Krummenacher feedback ($I_{\rm k}$) and the current in the comparator  
($I_{\rm DAC}$) are the other primary settings of the LIN FE. 

The current in the Krummenacher feedback system affects the duration of the signal discharge. Its default DAC value, KRUM\_CURR\_LIN = 29, is chosen so that a signal of \SI{6000}{e} results in an average of 6 ToT units, where one ToT unit corresponds to \SI{25}{ns}. 
For comparison, the most probable value for the signal generated in a non-irradiated  \SI{150}{\micro\meter} thick silicon sensor similar to the ones investigated in this paper was measured to be around \SI{11000}{e}, while 90$\%$ of the events have a signal below \SI{19000}{e}.
To study sensor performance, 
KRUM\_CURR\_LIN is reduced to 20 for measurements of the spatial resolution 
in order to increase the discharge time. This forces large signals originating from single-pixel clusters to concentrate in the last few ToT bins and provides more bins for smaller signals in multi-pixel clusters, hence improving the accuracy in the measurement of the charge sharing and ultimately the performance of the algorithm used to determine the hit position of a particle. It should be noted that longer ToT results in longer pixel deadtime which is a critical parameter at high hit rates. Therefore, for the operation of the inner IT layers a compromise between charge resolution and deadtime has to be found.

For all measurements at room temperature, the default value of LDAC\_LIN~=~130 was used. For measurements of irradiated modules at temperatures below \SI{0}{\celsius}, this value was increased to 185. This increases the output range of the per-pixel threshold adjustment and mitigates saturation of the trimming DAC owing to larger threshold dispersion. For the module irradiated to a fluence of $\Phi_{\rm eq} = $ \SI{2.1e16}{\per\centi\meter\squared}, LDAC\_LIN = 200 was used.

The global threshold, Vthreshold\_LIN, is set for each module individually to achieve the lowest possible threshold while keeping the Equivalent Noise Charge (ENC) below \SI{100}{e} and the number of noisy pixels below one percent after per-pixel adjustment. 
A pixel is considered noisy if it has a noise hit probability $>$\SI{2e-5}{} in a \SI{25}{ns} sensitive time window, which corresponds to 
2 hits in \SI{1e5}{} random triggers.
Per-pixel threshold adjustment is performed in two steps, using the charge injection system. 
First, the threshold of each pixel is adjusted using a series of iterative threshold scans at a high global threshold of approximately \SI{2500}{e}. In each iteration, the threshold of each pixel is compared to the mean of the thresholds of all pixels and its TDAC is adjusted in order to minimize the width of the threshold distribution. This results in a set of TDAC values for all pixels. Next, the ROC is configured using the TDAC values obtained in the first step and the trimming procedure is performed again at a lower global threshold of around 1000 e. This step produces a set of fine-tuned TDAC values for low-threshold operation.
The other parameters remained the same for all reported measurements except for the module irradiated to the highest fluence, where a different set of parameters was necessary as mentioned above (Table \ref{table_linfe}). 
%
%

\begin{table}[!t]
    \caption{The DAC settings for the LIN FE  used in the measurements of non-irradiated and irradiated modules. Different values of Vthreshold\_LIN were used for different modules, hence the ranges of DAC values are listed. The \SI{1}{\mega\electronvolt} neutron equivalent fluences $\Phi_{\rm eq}$ are in \SI{}{\per\centi\meter\squared} and total ionizing dose (TID) is in \SI{}{\mega\gray}. As explained in the text, in the measurements of the spatial resolution KRUM\_CURR\_LIN = 20 was used, while 29 was used in other measurements.}
\bigskip
\label{table_linfe}
\centering
\begin{tabular}{lccc}
    DAC name & $\Phi_{\rm eq}$ = \num{0} & \num{0 } < $ \Phi_{\rm eq}\le$ \num{1.0e16} & $\Phi_{\rm eq}$ =  \num{2.1e16}{} \\
     & TID = \num{0} & $\num{0 } < \text{ TID}\leq$ \num{13} & TID $\approx$ \num{30}{} \\ \hline\hline
    Vthreshold\_LIN    & 340 to 360  & 340 to 360  & 354        \\
    LDAC\_LIN          & 130         & 185         & 200        \\
    KRUMM\_CURR\_LIN   & 29 or 20    & 29 or 20    & 20         \\
    FC\_BIAS\_LIN      & 20          & 20          & 20         \\
    COMP\_LIN          & 110         & 110         & 110        \\
    PA\_IN\_BIAS\_LIN  & 350         & 350         & 250        \\
    REF\_KRUM\_LIN     & 300         & 300         & 300        \\ \hline
         
\end{tabular}
\end{table}

\begin{figure*}[!t]
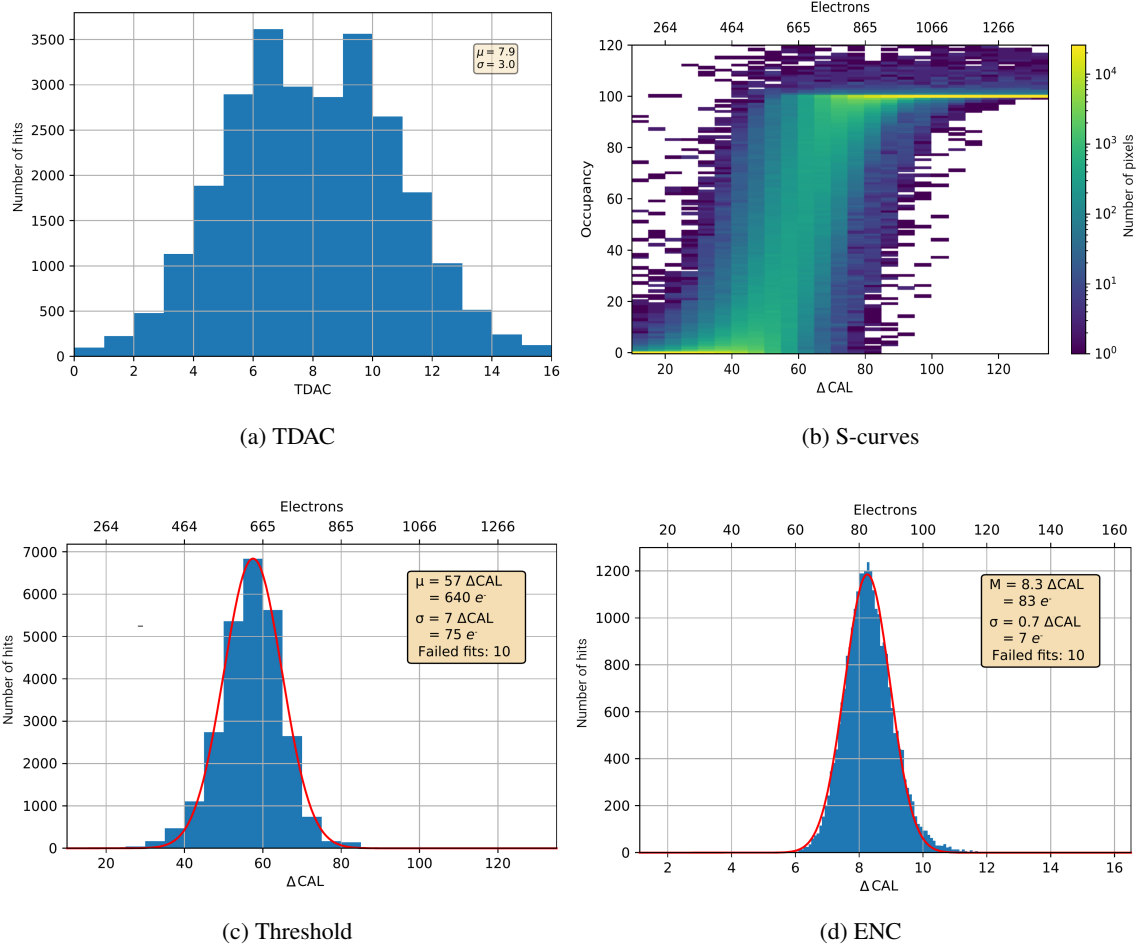

\centering
\subfloat[TDAC]{\includegraphics[width=0.49\linewidth]{images/m564_tdac.png}
\label{fig_m564-tdac}}
\hfil
\subfloat[S-curves]{\includegraphics[width=0.49\linewidth]{images/m564_scurve.png}
\label{fig_m564-scurve}}
\hfil
\subfloat[Threshold]{\includegraphics[width=0.49\linewidth]{images/m564_threshold.png}
\label{fig_m564-threshold}}
\hfil
\subfloat[ENC]{\includegraphics[width=0.49\linewidth]{images/m564_noise.png}
\label{fig_m564-noise}}
    \caption{(a) Distribution of TDAC, (b) S-curves for all pixels, (c) distribution of trimmed thresholds with Gaussian fit, and (d) ENC distribution  with Gaussian fit after tuning the ROC for a non-irradiated module with \SI{50x50}{\micro\meter} pixel size at room temperature and with a sensor bias voltage of \SI{120}{\volt}.}
\label{fig_m564}
\end{figure*}

The tuning results for a non-irradiated module with \SI{50x50}{\micro\meter} pixel size at room temperature and with a sensor bias voltage of \SI{120}{\volt} are shown in Fig.~\ref{fig_m564}. As visible in Fig.~\ref{fig_m564-tdac}, the distribution of the TDAC values after the tuning procedure for all pixels has an apparent double-peak structure, while a Gaussian distribution is expected.
This feature is not present anymore in the next version of the chip. 
The distribution has a mean of 7.9 and a standard deviation of 3.0. 
The threshold is determined using the charge injection system of the RD53A chip. A voltage step supplied to the charge injection capacitors determines the quantity of the injected charge. The amplitude of the voltage step, $\Delta\text{VCAL}$, is controlled by the difference of two 12-bit voltage DAC outputs. The corresponding difference in DAC units between the two DACs is referred to as $\Delta\text{CAL}$. Following the recommendation of the RD53A designers, the following conversion equation between $\Delta\text{CAL}$ and the charge Q in units of the elementary charge e is used:
\begin{equation}
\centering
    \text{Q(e)} = \SI{}{\Delta CAL} \cdot 10.02\,(\text{e}) + \SI{64}(\text{e}).
\end{equation}

The $\Delta\text{CAL}$ is scanned over a range of voltages and for each voltage the corresponding charge is injected into each pixel 100 times. The injections are performed according to a pattern so that charges 
are not injected into neighboring pixels at the same time. The pattern is then shifted to cover all pixels. After each pattern injection all of the pixels are read out. The pixel occupancy is defined as the total number of pixels above threshold (hits) divided by the total number of injections as:
\begin{equation}
\centering
    \text{Occupancy} = \frac{N_{\text{hit}}}{N_{\text{injection}}}.
\end{equation}
Plotting the occupancy for an individual pixel as a function of the amplitude of the injected charge ($\Delta\text{CAL}$) results in an S-curve, which is fitted with an error function. 
The value of $\Delta\text{CAL}$ at 50\% occupancy determines the threshold of an individual pixel. 
Figure~\ref{fig_m564-scurve} shows the overlay of S-curves for all \emph{Linear front-end} pixels in a single ROC, 
where the number of pixels in a particular bin is indicated by the color-scale.
The distribution of the pixel thresholds is shown in Fig.~\ref{fig_m564-threshold}. 
%
%
The mean threshold for this module is \SI{640}{e} with a variation between the pixels given by a standard deviation of \SI{75}{e}. 
A fit is reported as failed if the  $\Delta\text{VCAL}$-range selected for the threshold scan is too small, if  
$\chi^2/\text{DOF}>40$, or if the occupancy does not increase with $\Delta\text{VCAL}$ for a given pixel. The number of failed fits for this module is negligible.  
The sigma of the error function is taken to be the ENC. The module has a mean ENC  of \SI{83}{e} with a standard deviation of  \SI{7}{e}, as shown in Fig.~\ref{fig_m564-noise}. 
%

Figure~\ref{fig_m521} illustrates a similar set of plots for a module irradiated with protons at the MC40 cyclotron facility in Birmingham to a fluence of 
$\Phi_{\rm eq }$ = \SI{ 5.4e15}{\per\centi\meter\squared}.  
This corresponds to a total ionizing dose of \SI{7.7}{\mega\gray} in silicon dioxide, relevant for surface damage in the sensor. The tuning was performed for a sensor temperature of approximately  \SI{-30}{\celsius} with a sensor bias voltage of \SI{200}{\volt}. 
A significant number of pixels have TDAC values of 0 or 15, a sign that the  dynamic range of the TDAC circuit is not sufficient to compensate for the threshold dispersion after irradiation.
The mean threshold has increased significantly to 1230~e, and the threshold and ENC distributions have larger spreads after irradiation: \SI{144}{e} and \SI{11}{e}, respectively. The number of failed fits has increased to 97, which is in the range of  0.4\% of all pixels in the \emph{Linear front-end}.

\begin{figure*}[!t]
\centering
\subfloat[TDAC]{\includegraphics[width=0.49\linewidth]{images/m521_tdac.png}
\label{fig_m521-tdac}}
\hfil
\subfloat[S-curves]{\includegraphics[width=0.49\linewidth]{images/m521_scurve.png}
\label{fig_m521-scurve}}
\hfil
\subfloat[Threshold]{\includegraphics[width=0.49\linewidth]{images/m521_threshold.png}
\label{fig_m521-threshold}}
\hfil
\subfloat[ENC]{\includegraphics[width=0.485\linewidth]{images/m521_noise.png}
\label{fig_m521-noise}}
\caption{(a) Distribution of TDAC, (b) S-curves for all pixels, (c) distribution of trimmed thresholds with Gaussian fit, and (d) ENC distribution  with Gaussian fit after tuning the ROC for a module with \SI{50x50}{\micro\meter} pixel size irradiated to a fluence of $\Phi_{\rm eq} = $ \SI{5.3e15}{\per\centi\meter\squared}. The tuning was performed at approximately \SI{-30}{\celsius} with a sensor bias voltage of \SI{200}{\volt}. 
The TDAC distribution shows a significant number of entries in the last bin 
because of the saturation of the dynamic range of the TDAC circuit at low temperatures.  
In the next generation of the chip, this issue has been addressed by adding an additional bit to the TDAC circuit.
The entries with low occupancy  at high injection voltages in the S-curve plot are likely associated with the combination of this feature and the accumulated irradiation damage.
}
\label{fig_m521}
\end{figure*}

\section{Crosstalk}\label{xtalk}
\begin{figure*}[!t]
\centering
\subfloat[]{\includegraphics[width=0.49\linewidth]{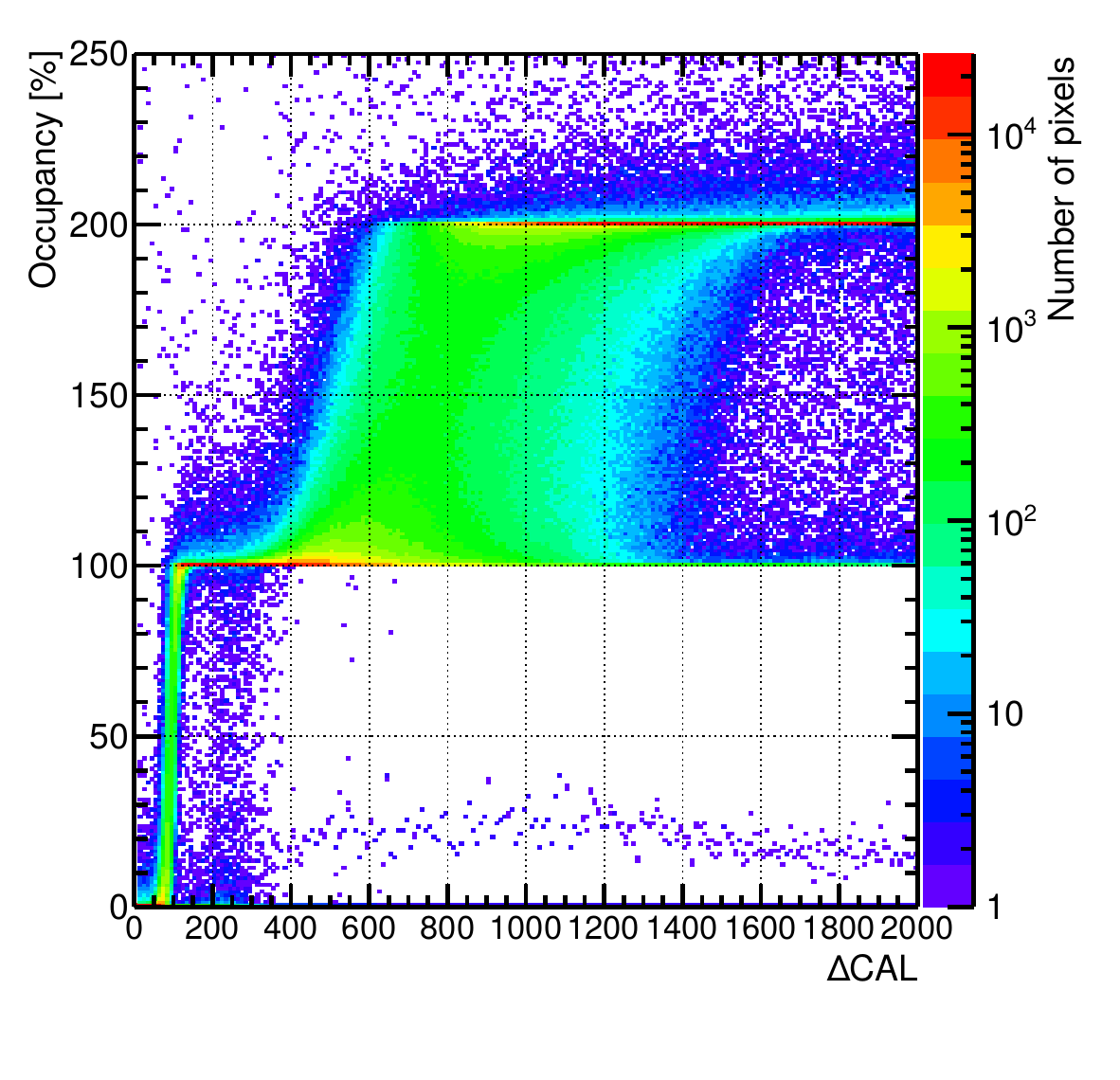}
\label{fig_m529-xtscurve}}
\hfil
\subfloat[]{\includegraphics[width=0.49\linewidth]{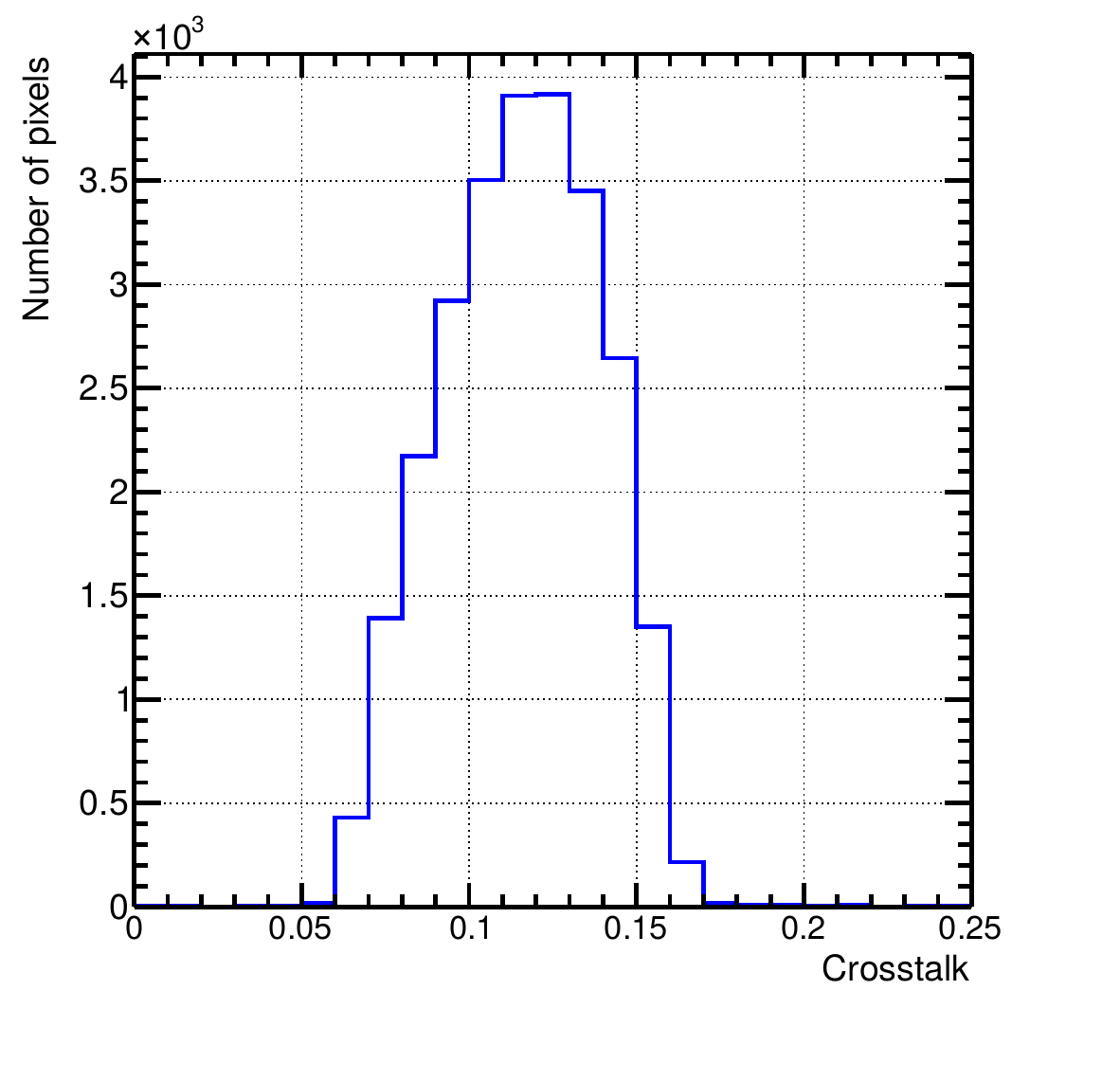}
    \label{fig_m529-xt}}
    \caption{(a) 
    Occupancy versus $\Delta\text{CAL}$ for a non-irradiated RD53A module with a sensor with pixel cell size of \SI{100x25}{\micro\meter}. 
    Crosstalk leads to a double S-curve. (b) The crosstalk distribution for all pixels. An average crosstalk of approximately 11\% is observed in these modules. The measurements were performed at room temperature and a bias voltage of \SI{120}{\volt}.}
\label{fig_xt}
\end{figure*}

To be able to bump bond a sensor with a pixel pitch of \SI{100x25}{\micro\meter} to the RD53A chip, which has a bump bond pattern of \SI{50x50}{\micro\meter}, the bump bonding pads on the sensor are placed on the boundary between two neighboring pixels, as shown in Fig.~\ref{fig_rd53a-100}. These pads overlap with the implant layer of the neighboring pixels, and as a result crosstalk between the two pixels via capacitive coupling is observed.
%
The amount of crosstalk is determined from S-curve scans using the charge injection system of the RD53A chip as described in Sec.~\ref{tuning}.
In the absence of crosstalk, the maximum pixel occupancy should remain at 100\% regardless of the injection amplitude. 
In case of crosstalk, a charge injected into a single pixel leads to a charge above threshold also in the neighbor pixel, provided that the injected charge is large enough.
The plot of occupancy versus $\Delta\text{CAL}$ therefore shows a second S-curve leading to an occupancy 
of up to 200\%. 
The point where the mean occupancy is 150\%, 
$\Delta\rm CAL_{150}$, is the amount of charge injected in units of $\Delta\text{CAL}$ that leads to half of the injections resulting in the signal of a neighbor pixel going above threshold.
As seen in Fig.~\ref{fig_m529-xtscurve} for a sensor with \SI{100x25}{\micro\meter} pixel size, this kind of double S-curve is present. In sensors with \SI{50x50}{\micro\meter} pixel size the occupancy remains at 100\% even at very high injection amplitudes.

The double S-curve of each pixel is fitted using error functions to determine the ratio of the $\Delta\text{CAL}$ at 50\% occupancy ($\Delta\rm CAL_{50}$) to the $\Delta\text{CAL}$ at 150\% occupancy ($\Delta\rm CAL_{150}$) for each pixel:

\begin{equation}
\centering
    \text{Crosstalk} = \frac{\Delta \rm CAL_{50}}{\Delta \rm CAL_{150}}.
\end{equation}
The distribution of crosstalk for all pixels at a global threshold of \SI{900}{e} is shown in Fig.~\ref{fig_m529-xt}. The mean crosstalk is  approximately 11\%. TCAD simulations show that a semi-circular cutout in the implant just under the bump bonding pads reduces the capacitive coupling to the neighbor pixel from about \SI{21} to \SI{14}{fF}, 
thereby reducing the crosstalk. Such a design has been implemented in later production submissions~\cite{Moh}.

\section{Sensor Evaluation with Particle Beams}\label{testbeam}
\begin{figure*}[!b]
\centering
\subfloat[]{

    \resizebox{\linewidth}{!}{%
        \tikzset{every picture/.style={line width=0.75pt}} 

    \begin{tikzpicture}[x=0.75pt,y=0.75pt,yscale=-1,xscale=1]

        \draw [line width=2.25]    (151.5,89) -- (152,201) ;
        \draw [line width=2.25]    (201.5,89) -- (202,201) ;
        \draw [line width=2.25]    (251.5,89) -- (252,201) ;
        \draw [line width=2.25]    (381.5,89) -- (382,201) ;
        \draw [line width=2.25]    (431.5,89) -- (432,201) ;
        \draw [line width=2.25]    (481.5,89) -- (482,201) ;
        \draw [color={rgb, 255:red, 208; green, 2; blue, 27 }  ,draw opacity=1 ][line width=1.5]    (644.5,143) -- (56.5,145.98) ;
        \draw [shift={(53.5,146)}, rotate = 359.71000000000004] [color={rgb, 255:red, 208; green, 2; blue, 27 }  ,draw opacity=1 ][line width=1.5]    (14.21,-4.28) .. controls (9.04,-1.82) and (4.3,-0.39) .. (0,0) .. controls (4.3,0.39) and (9.04,1.82) .. (14.21,4.28)   ;
        \draw [color={rgb, 255:red, 189; green, 16; blue, 224 }  ,draw opacity=1 ][line width=3]    (524.5,124) -- (525,166) ;
        \draw [color={rgb, 255:red, 65; green, 117; blue, 5 }  ,draw opacity=1 ][line width=3]    (315.5,124) -- (316,166) ;
        \draw [color={rgb, 255:red, 100; green, 100; blue, 100 }  ,draw opacity=1 ][line width=1.5]  (108.4,226) -- (108.4,170)(58,220.4) -- (114,220.4) (103.4,177) -- (108.4,170) -- (113.4,177) (65,225.4) -- (58,220.4) -- (65,215.4)  ;
        \draw  [color={rgb, 255:red, 100; green, 100; blue, 100 }  ,draw opacity=1 ][line width=1.5]  (152,213) .. controls (152,217.67) and (154.33,220) .. (159,220) -- (192,220) .. controls (198.67,220) and (202,222.33) .. (202,227) .. controls (202,222.33) and (205.33,220) .. (212,220)(209,220) -- (245,220) .. controls (249.67,220) and (252,217.67) .. (252,213) ;
        \draw  [color={rgb, 255:red, 100; green, 100; blue, 100 }  ,draw opacity=1 ][line width=1.5]  (382,213) .. controls (382,217.67) and (384.33,220) .. (389,220) -- (422,220) .. controls (428.67,220) and (432,222.33) .. (432,227) .. controls (432,222.33) and (435.33,220) .. (442,220)(439,220) -- (475,220) .. controls (479.67,220) and (482,217.67) .. (482,213) ;
        \draw [color={rgb, 255:red, 74; green, 144; blue, 226 }  ,draw opacity=1 ][line width=2.25]    (587.5,118) -- (620.5,172) ;

        \draw (145,231) node [anchor=north west][inner sep=0.75pt]  [color={rgb, 255:red, 100; green, 100; blue, 100 }  ,opacity=1 ] [align=left] {Downstream arm};
        \draw (385,231) node [anchor=north west][inner sep=0.75pt]  [color={rgb, 255:red, 100; green, 100; blue, 100 }  ,opacity=1 ] [align=left] {Upstream arm};
        \draw (66,224) node [anchor=north west][inner sep=0.75pt]  [color={rgb, 255:red, 100; green, 100; blue, 100 }  ,opacity=1 ] [align=left] {z};
        \draw (113,178) node [anchor=north west][inner sep=0.75pt]  [color={rgb, 255:red, 100; green, 100; blue, 100 }  ,opacity=1 ] [align=left] {y};
        \draw (70,127) node [anchor=north west][inner sep=0.75pt]   [align=left] {\textbf{\textcolor[rgb]{0.82,0.01,0.11}{$e^{-}$}}};
        \draw (300,173) node [anchor=north west][inner sep=0.75pt]  [color={rgb, 255:red, 100; green, 100; blue, 100 }  ,opacity=1 ] [align=left] {DUT};
        \draw (489,173) node [anchor=north west][inner sep=0.75pt]  [color={rgb, 255:red, 100; green, 100; blue, 100 }  ,opacity=1 ] [align=left] {Scintillator};
        \draw (570,174) node [anchor=north west][inner sep=0.75pt]  [color={rgb, 255:red, 100; green, 100; blue, 100 }  ,opacity=1 ] [align=left] {Timing ref.};

    \end{tikzpicture}}\label{fig_tele}}

\label{fig_tb}
\hfil
\subfloat[]{\includegraphics[width=\linewidth]{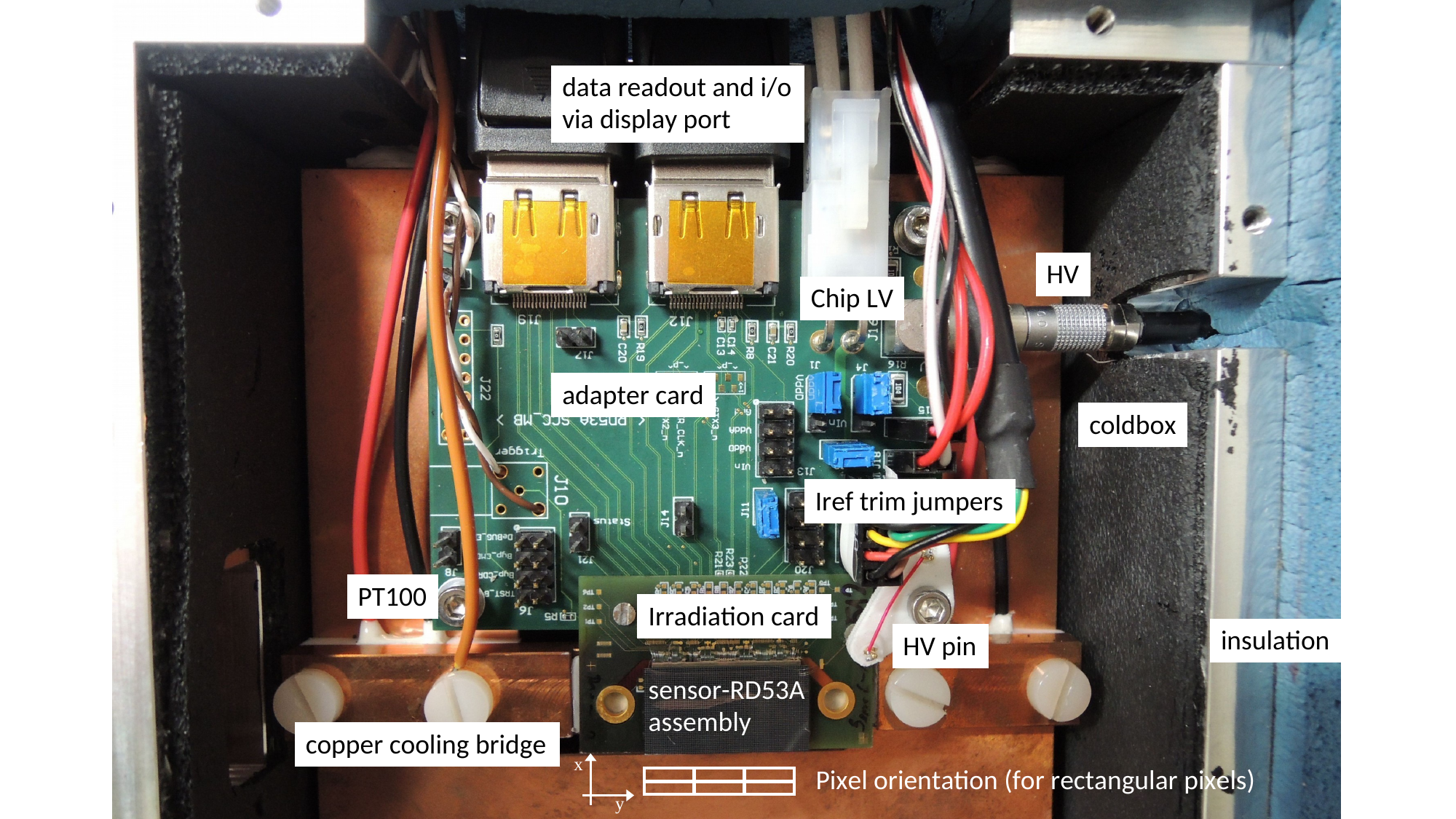}
\label{fig_coldbox}}
\caption{(a) Schematic side view of the testbeam setup with the global testbeam and telescope coordinate system. (b) A RD53A module mounted inside the coldbox.}
\end{figure*}
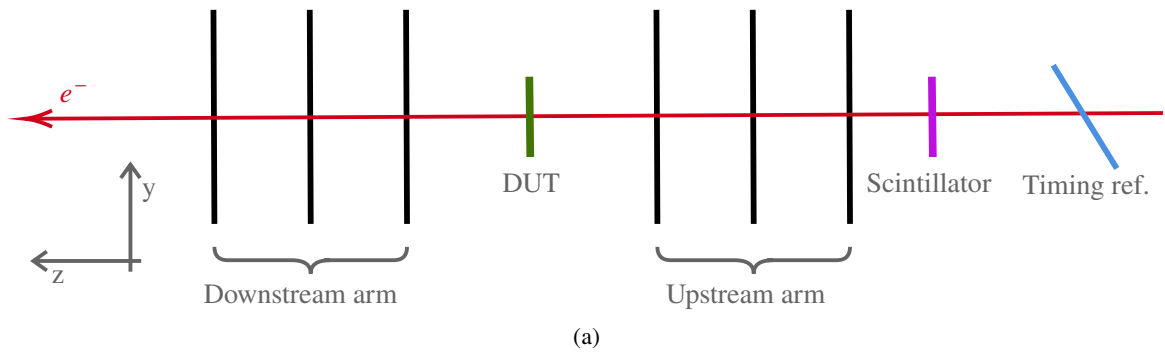
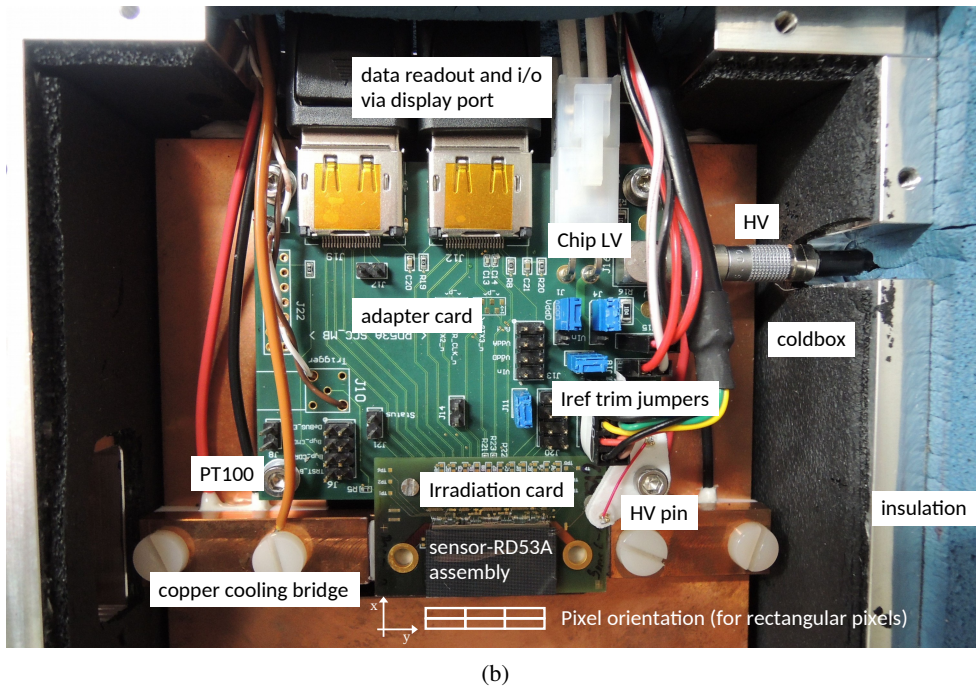

The performance of the pixel modules has been evaluated using electron beams at the DESY-II testbeam facility~\cite{desytb}. DESY-II is an electron/positron synchrotron that offers secondary beams with user-selectable momenta from \SI{1} to \SI[per-mode=symbol]{6}{\giga\electronvolt\per\clight}. The data for the studies presented in this paper were taken with electron beams of \SI[per-mode=symbol]{3}{\giga\electronvolt\per\clight}.

The testbeam facility is equipped with EUDET-type pixel beam telescopes~\cite{eudet} for particle tracking. The DATURA beam telescope at beam line 21 and the AZALEA beam telescope at beam line 24 were used in the following studies. A schematic side view of the setup at the beam line is shown in Fig.~\ref{fig_tele}. Each telescope consists of six planes, divided into an upstream arm and a downstream arm, with three planes each. The device under test (DUT) was mounted on a movable stage between the two arms.
The integration time of the \SI{50}{\micro\meter} thick MIMOSA-26 chips~\cite{mimosa-26} used in the telescope is \SI{115.2}{\micro\second}, corresponding to 115 turns of the DESY-II synchrotron. Owing to such a long integration time, multiple tracks are present in each telescope event, referred to as pileup. A CMS Phase-1 pixel module (\SI{150x100}{\micro\meter} pixel size)~\cite{Adam_2021} with an integration time of \SI{25}{ns} was placed in front of the upstream arm as a timing reference to reduce pileup of telescope tracks in the analysis. The reference module was inclined around the $x$ and $y$ axes to improve its spatial resolution.

Non-irradiated modules were mounted on an aluminium frame and measured at the ambient temperature of the experimental hall. Irradiated modules were mounted inside a thermally insulated coldbox,  which kept the modules at low temperatures in order to reduce the leakage current (Fig.~\ref{fig_coldbox}). The box was cooled down using a chiller and two Peltier elements. The chiller was set to a temperature of \SI{-25}{\celsius}. The box was continuously flushed with dry air to prevent condensation. Under these conditions, the temperature of the sensor is estimated to be approximately \SI{-30}{\celsius}.
The support for the DUT included a rotation stage with the axis of rotation pointing upward (in the telescope $y$ direction). 
For sensor modules with \SI{100x25}{\micro\meter} pixel size, the 
entire support frame or cold box was rotated by \SI{90}{\degree} with respect to the orientation shown in Fig.~\ref{fig_coldbox} when rotations around the long pixel side were desired.
Independent of the orientation of the DUT, resolution plots always refer to local sensor coordinates where $x$ is the coordinate parallel to the \SI{25}{\micro\meter} 
pixel side. 
%
The telescope, DUT, and reference module were synchronized using a common trigger signal, which was generated by the coincidence of a pair of trigger scintillators placed upstream of the first telescope plane.

\subsection{Data Analysis}
In the first step of the data analysis, tracks in the six telescope planes are reconstructed and used to determine the relative positioning of the planes (telescope alignment). Since the track reconstruction requires alignment and alignment requires reconstructed tracks, the process starts using loose cuts on the maximum spatial distance between clusters in the planes and a track and is refined in an iterative manner to achieve tighter cuts and improved alignment. In each iteration the position and rotation of each telescpe plane is corrected using the parameters of the reconstructed tracks.

In the next step, telescope tracks are extrapolated to and checked against the hits in the reference module to remove out-of-time tracks. Tracks within a distance of \SI{150}{\micro\meter} in $x$ and \SI{100}{\micro\meter} in $y$ from a hit in the reference module are accepted and are used to determine the relative alignment of the reference module. These tracks are also used in any subsequent analysis.

The last step involves performing the alignment of the DUT by extrapolating the tracks to the position of the DUT. Owing to the presence of a copper bar behind the DUT in the coldbox, the downstream track resolution is significantly deteriorated for data taken with irradiated modules. Therefore, only the upstream triplet is used for the alignment and analysis in this case. For modules with \SI{100x25}{\micro\meter} pixel size, hits in the DUT within a distance  from a track of \SI{150}{\micro\meter} in the long pixel direction and \SI{100}{\micro\meter} in the short pixel direction are assigned to the track. For modules with \SI{50x50}{\micro\meter} pixel size, hits within a \SI{100}{\micro\meter} distance in both directions are accepted. A hit is defined as a pixel with a signal above threshold.
Adjacent hit pixels are combined into a cluster and its position in local coordinates is calculated as the weighted mean of the 
pixel positions with the pixel charges in units of ToT as weights (center-of-gravity reconstruction method). The difference between the extrapolation of the track and the position of the cluster in the DUT is used to determine the alignment of the DUT with respect to the telescope. Six alignment parameters are determined for the DUT: three local position coordinates and three rotation angles.

\subsection{Hit Efficiency}
\begin{figure}[!t]
\centering
\includegraphics[width=0.75\linewidth]{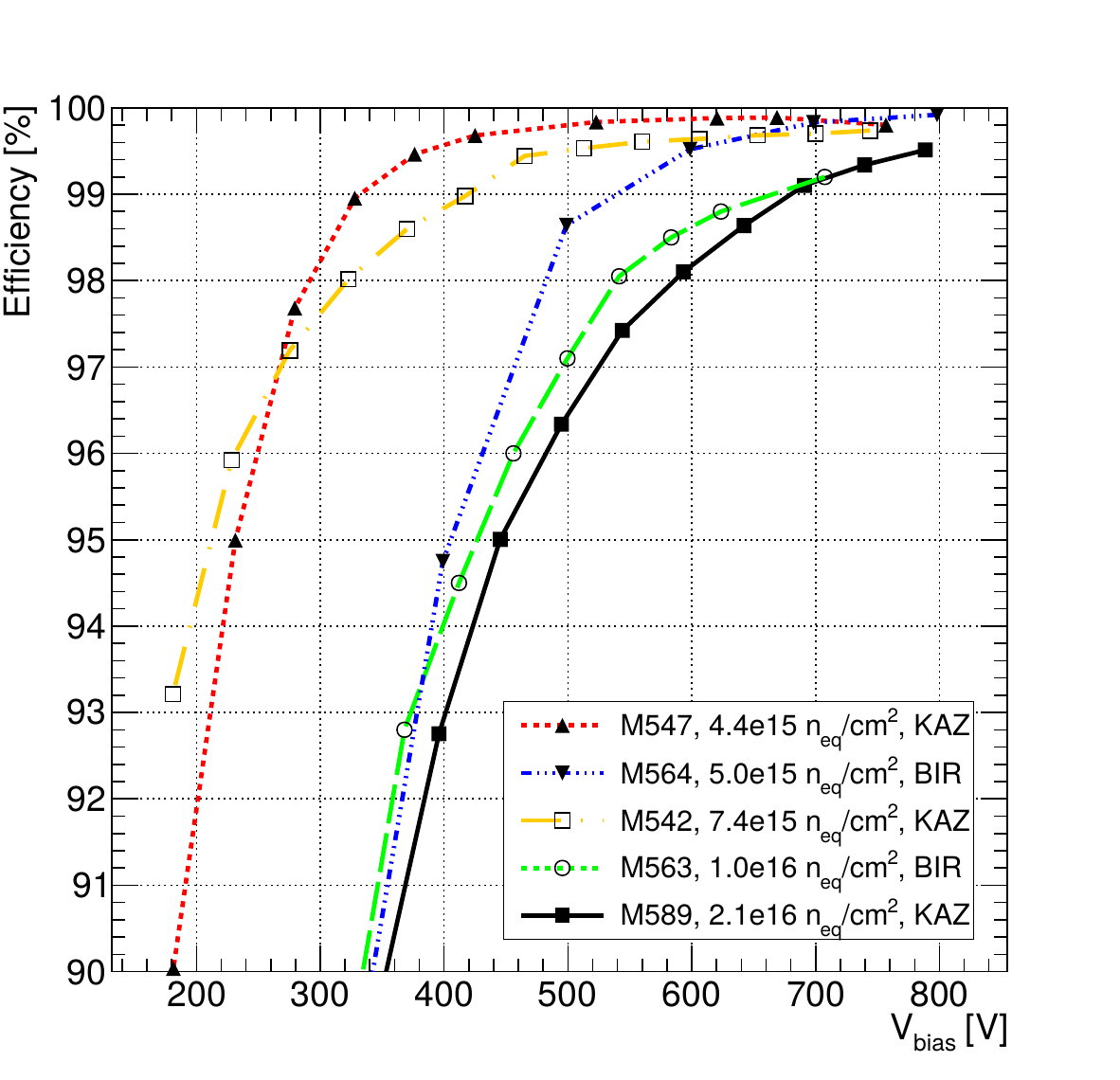}
    \caption{Efficiency as a function of bias voltage for RD53A sensor modules without a bias dot after irradiation with protons at the Karlsruhe Cyclotron (KAZ) and at the MC40 cyclotron facility in Birmingham (BIR). All modules have been measured at a temperature below \SI{-25}{\celsius} at vertical incidence and achieve an efficiency in excess of 99\%.}
\label{fig_compEff}
\end{figure}

The hit efficiency is studied for non-irradiated and irradiated pixel modules. It is defined as:
\begin{equation}
\centering
    \epsilon_{\rm{hit}} = \frac{N_{\rm hit}}{N_{\rm t}},
\end{equation}
where $N_{\rm hit}$ is the number of reconstructed in-time tracks with a hit on the DUT within an acceptance region that corresponds to the 
\emph{Linear front-end}, excluding its two last rows and columns, 
and $N_{\rm t}$ is the total number of tracks traversing the DUT.

Non-irradiated modules without a bias dot achieve an efficiency greater than 99\% at bias voltages as low as \SI{10}{\volt}, which is well below the full depletion voltage of $\approx$\SI{75}{\volt}.

Modules without a bias dot irradiated to fluences of $\Phi_{\rm eq} = $ \SI{4.4}, \SI{5.4}, \SI{7.4}, \SI{10}, and \SI{21e15}{\per\centi\meter\squared} were investigated. As shown in Fig.~\ref{fig_compEff}, all these modules reach 99\% hit efficiency. Although significantly higher bias voltages are required to reach the efficiency goal, these voltages are still below the \SI{800}{\volt} limit considered for the operation in CMS. 
It is not understood why the efficiency curve for M542 is 
systematically above the one for M564 despite the larger fluence.

\begin{figure*}[!t]
\centering
\subfloat[]{\includegraphics[width=0.49\linewidth]{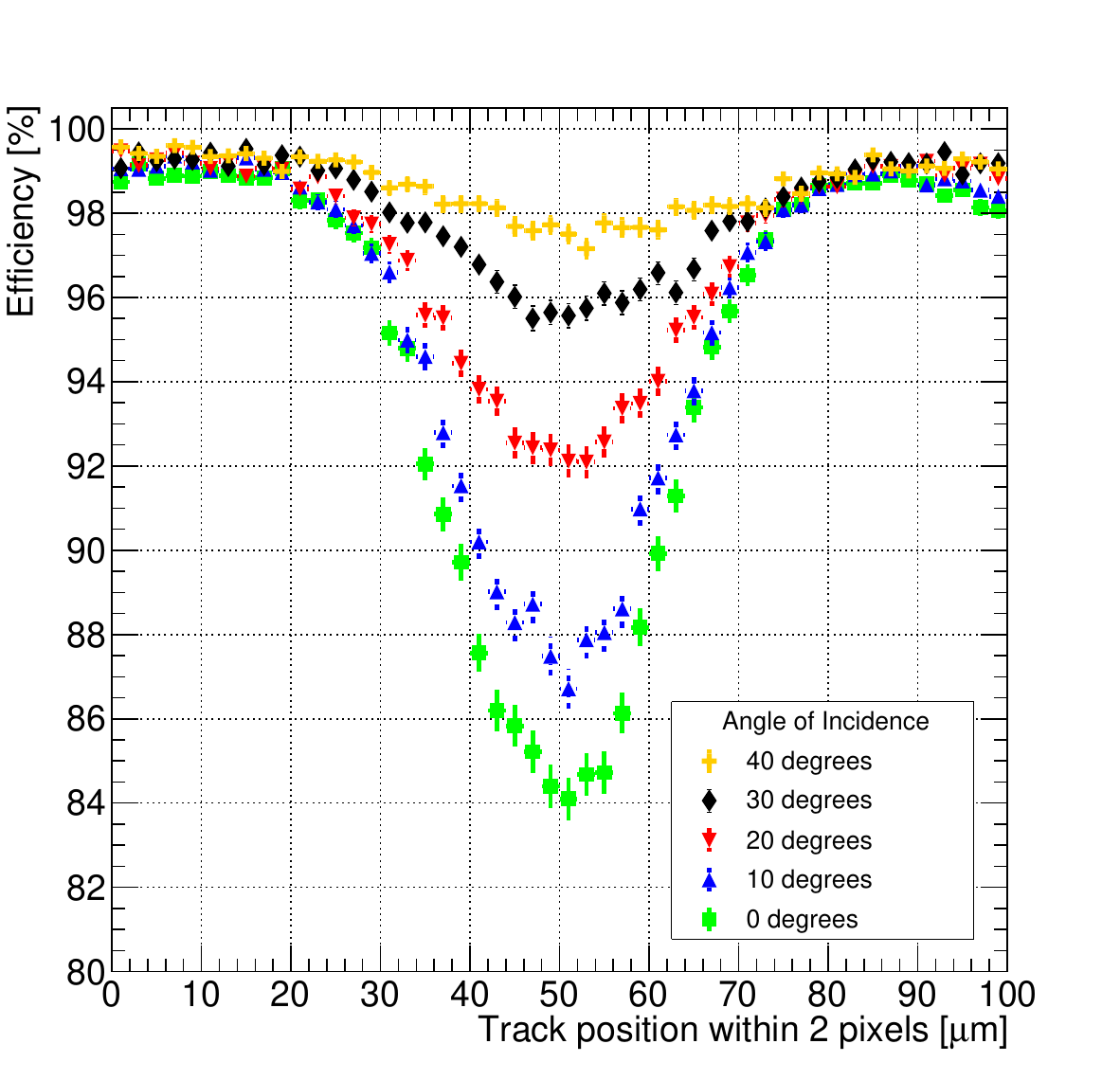}
\label{fig_m521-eff2}}
\hfil
\subfloat[]{\includegraphics[width=0.49\linewidth]{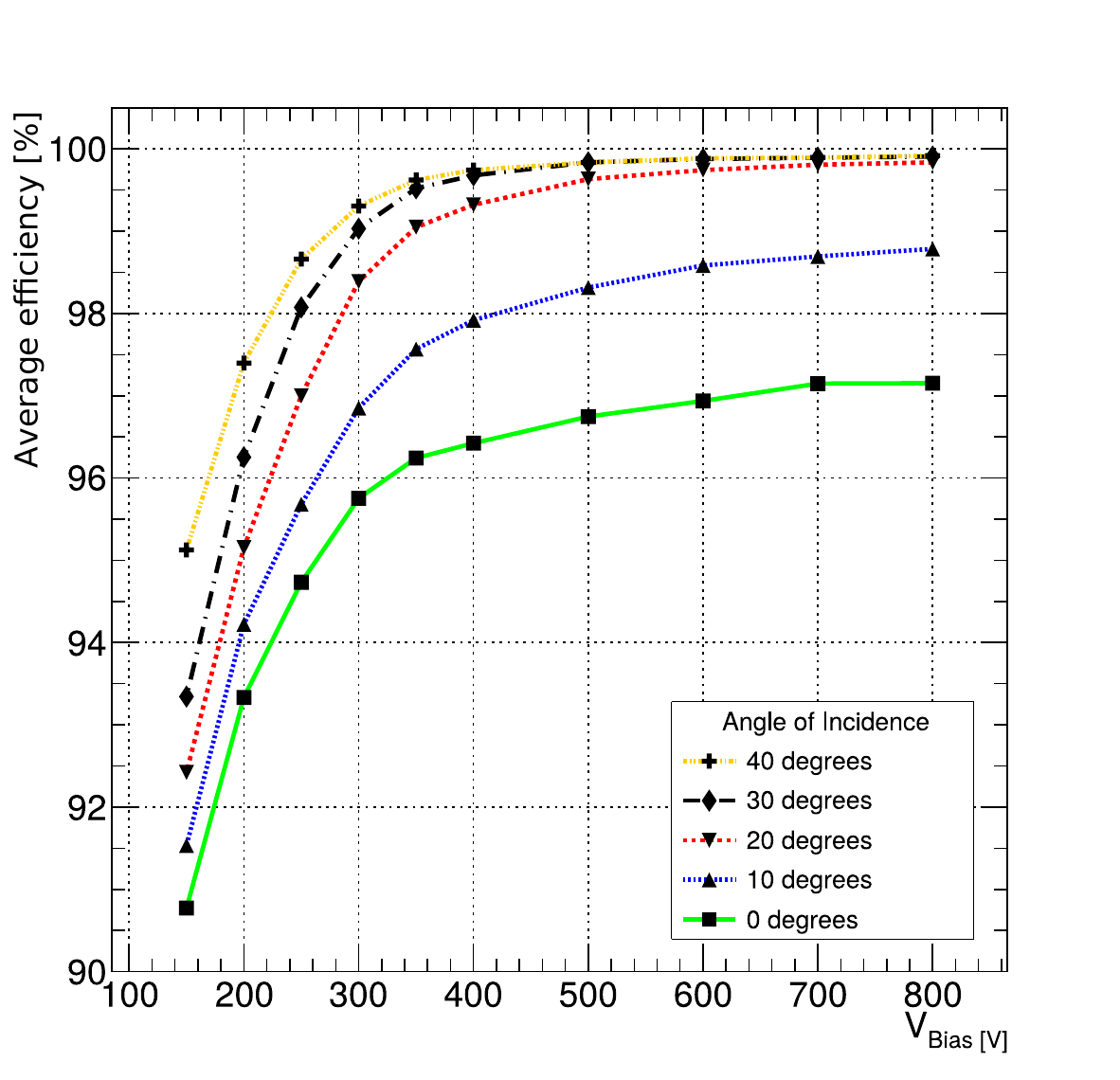}
\label{fig_m521-eff1}}
    \caption{(a) Efficiency as function of position in a \SI{2x2}{} pixel cell for a \SI{50x50}{\micro\meter} pitch sensor with a punch-through bias dot irradiated to $\Phi_{\rm eq} = $ \SI{5.3e15}{\per\centi\meter\squared} for different incident angles and measured at a bias voltage of \SI{250}{\volt}. (b) Efficiency as a function of bias voltage for the same module for a variety of incident angles. Irradiated modules are measured at a temperature below \SI{-25}{\celsius}.
    These data are also published in Ref.~\cite{cms:sensor2}.
    }
\label{fig_m521-eff}
\end{figure*}

Modules with a punch-through bias dot suffer from charge loss when a particle hits the bias dot at a small angle with respect to the module normal. Hence, a hit inefficiency in the vicinity of the bias dot is expected, as shown in Fig.~\ref{fig_m521-eff2}, which displays the efficiency as a function of impact position in a \SI{2x2}{} pixel cell at a bias voltage of \SI{250}{\volt}.
A voltage well below full depletion was chosen for this study to enhance the effect.
This is a module with \SI{50x50}{\micro\meter} pixel size irradiated to a fluence of $\Phi_{\rm eq} = $ \SI{5.3e15}{\per\centi\meter\squared}. 
The pointing resolution of the beam telescope at the DUT is on the order of \SI{8}{\micro\meter} for this measurement.
The efficiency as a function of bias voltage for different angles of incidence is shown in Fig.~\ref{fig_m521-eff1}. An efficiency in excess of 99\% is achieved for bias voltages above \SI{350}{\volt} for incident angles above \SI{20}{\degree}.
Based on the studies presented in this section, CMS has chosen sensors without punch-through bias dots for the 
upgrade of the IT.

\subsection{Spatial Resolution}
\begin{figure}[!t]
\centering
\includegraphics[width=0.75\linewidth]{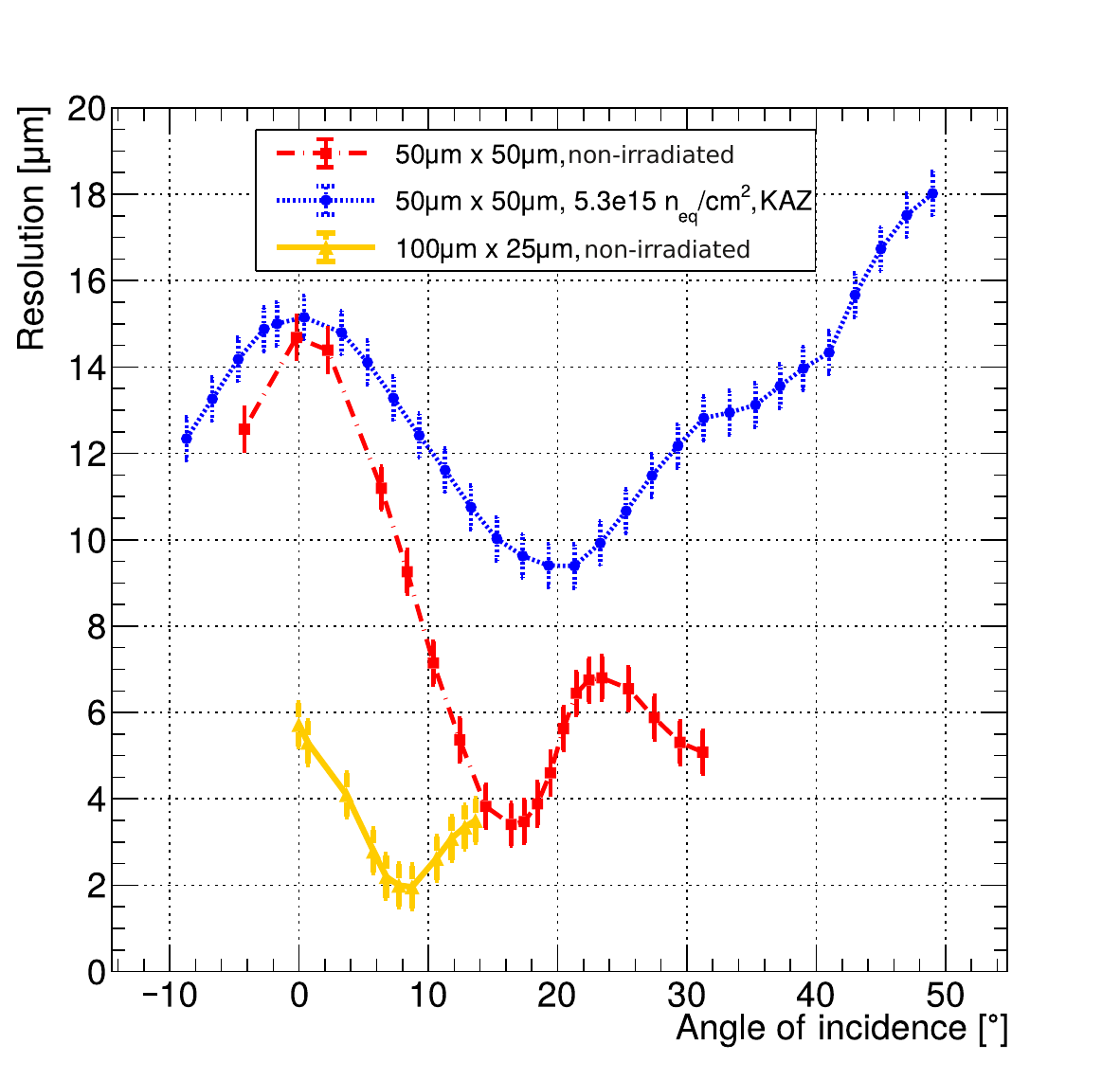}
    \caption{Spatial resolution as a function of incidence angle for RD53A modules equipped with a \SI{50x50}{\micro\meter} or \SI{100x25}{\micro\meter} sensor. 
    The modules are rotated around an axis that is parallel to the \SI{100}{\micro\meter} pixel side for sensors with rectangular pixels, and the resolution is determined in the coordinate parallel to the \SI{25}{\micro\meter} side.
    The non-irradiated modules were measured at room temperature at a bias voltage of \SI{120}{\volt}. The irradiated module was measured at a temperature below \SI{-25}{\celsius} and a bias voltage of \SI{800}{\volt}. The optimum angle for charge sharing is clearly visible for non-irradiated and irradiated sensors.
    }
\label{fig_compRes}
\end{figure}
\begin{figure}[!b]
\centering

\includegraphics[width=0.4\linewidth]{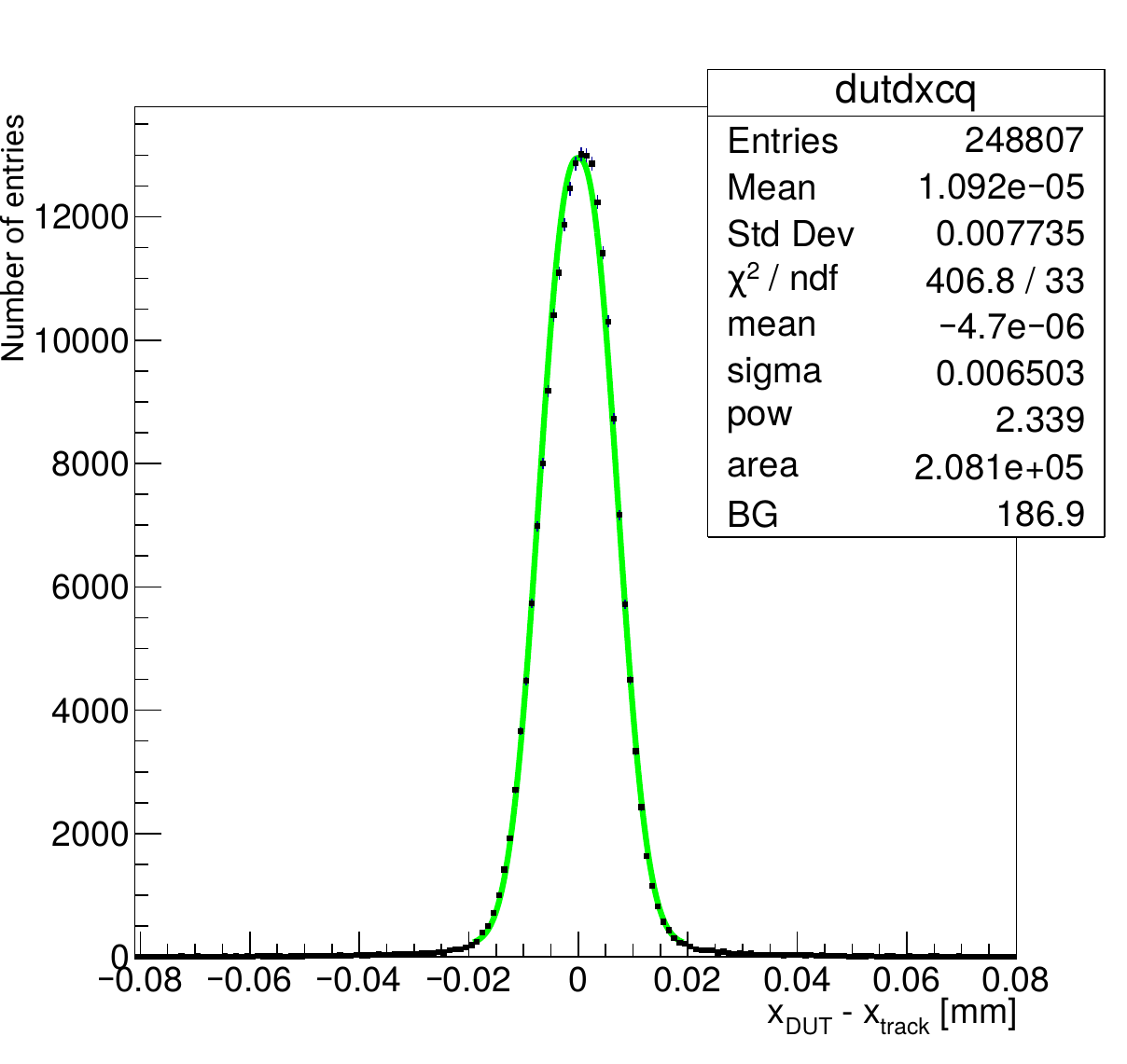}
\includegraphics[width=0.4\linewidth]{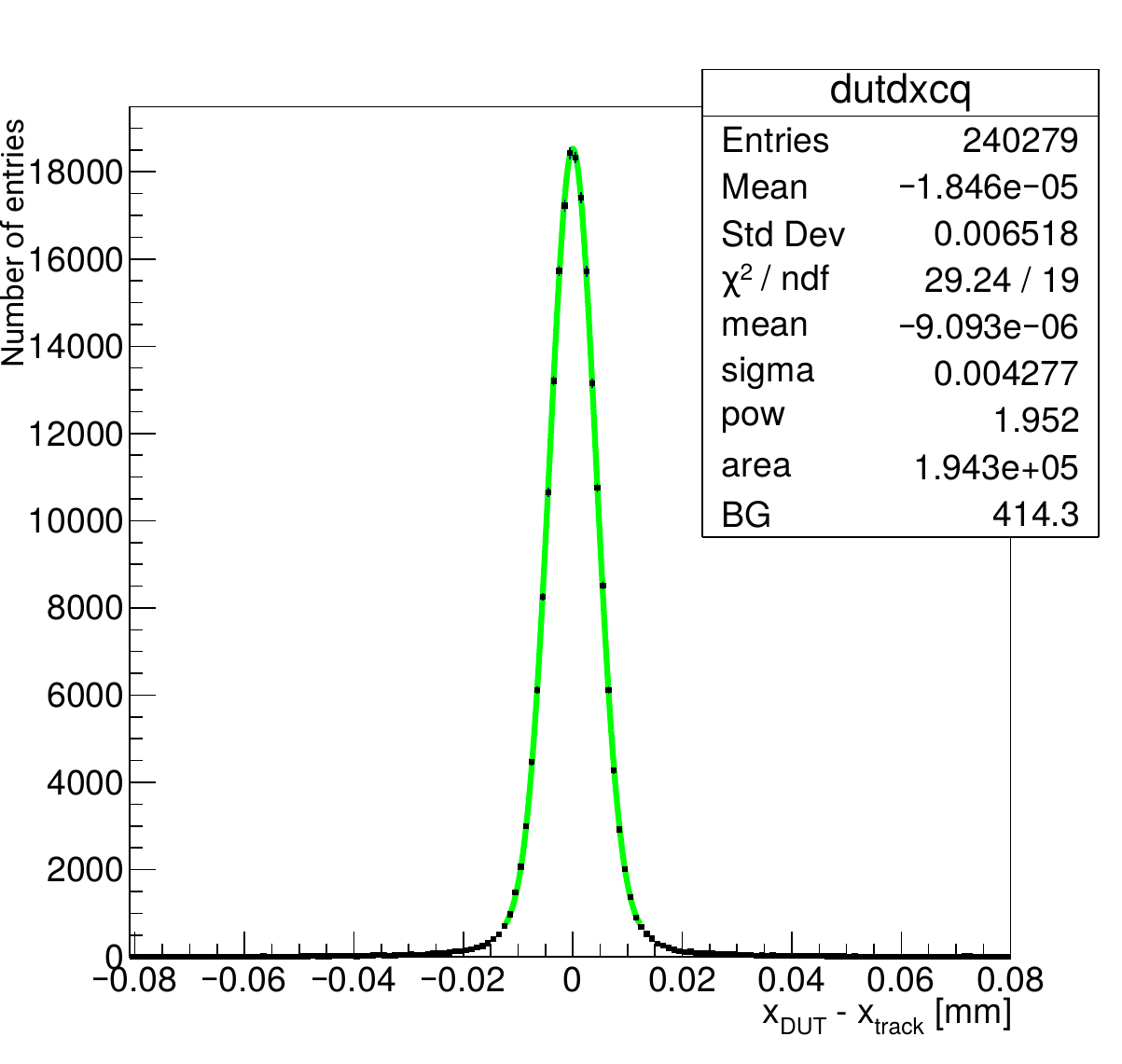}
\includegraphics[width=0.4\linewidth]{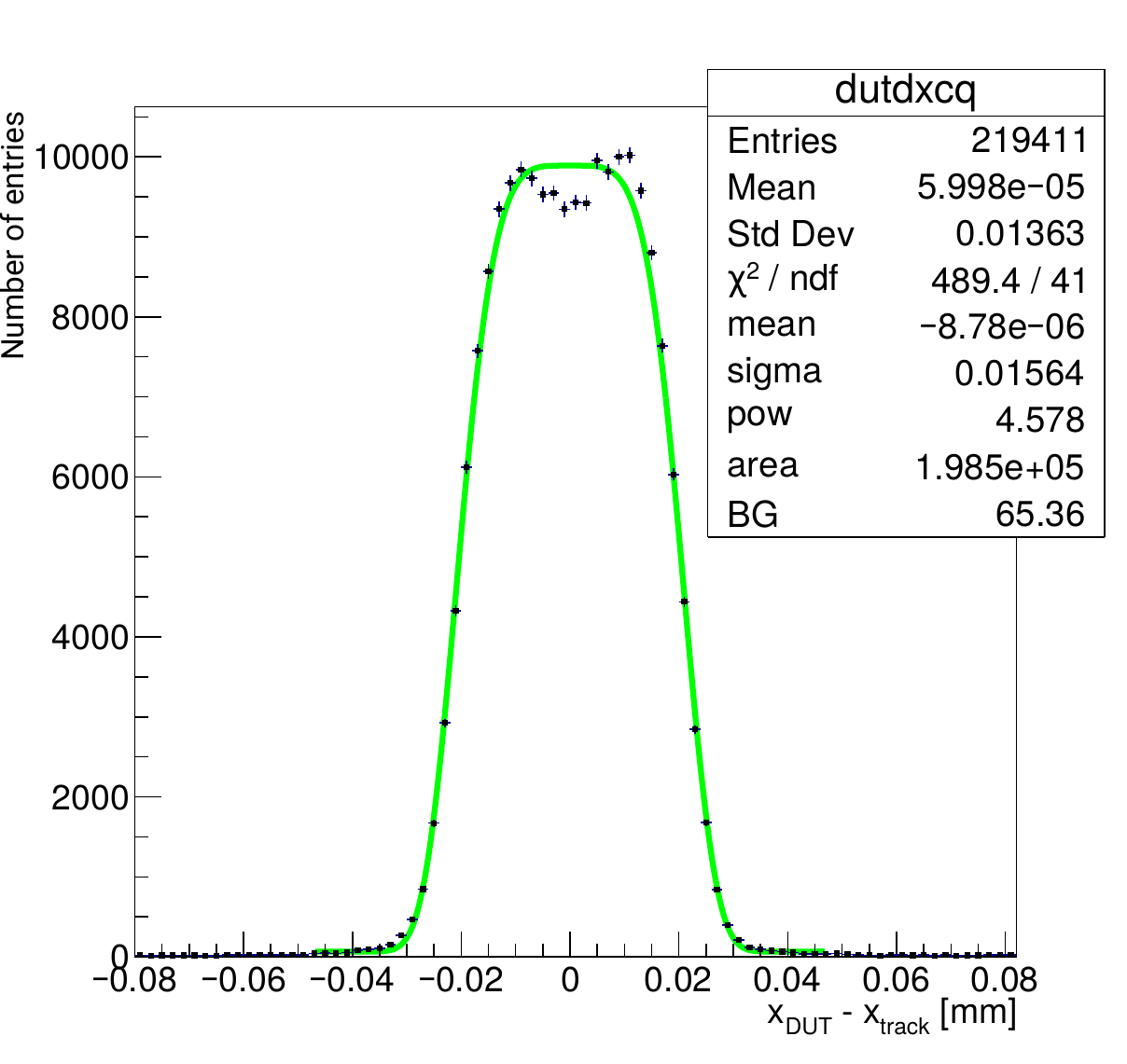}
\includegraphics[width=0.4\linewidth]{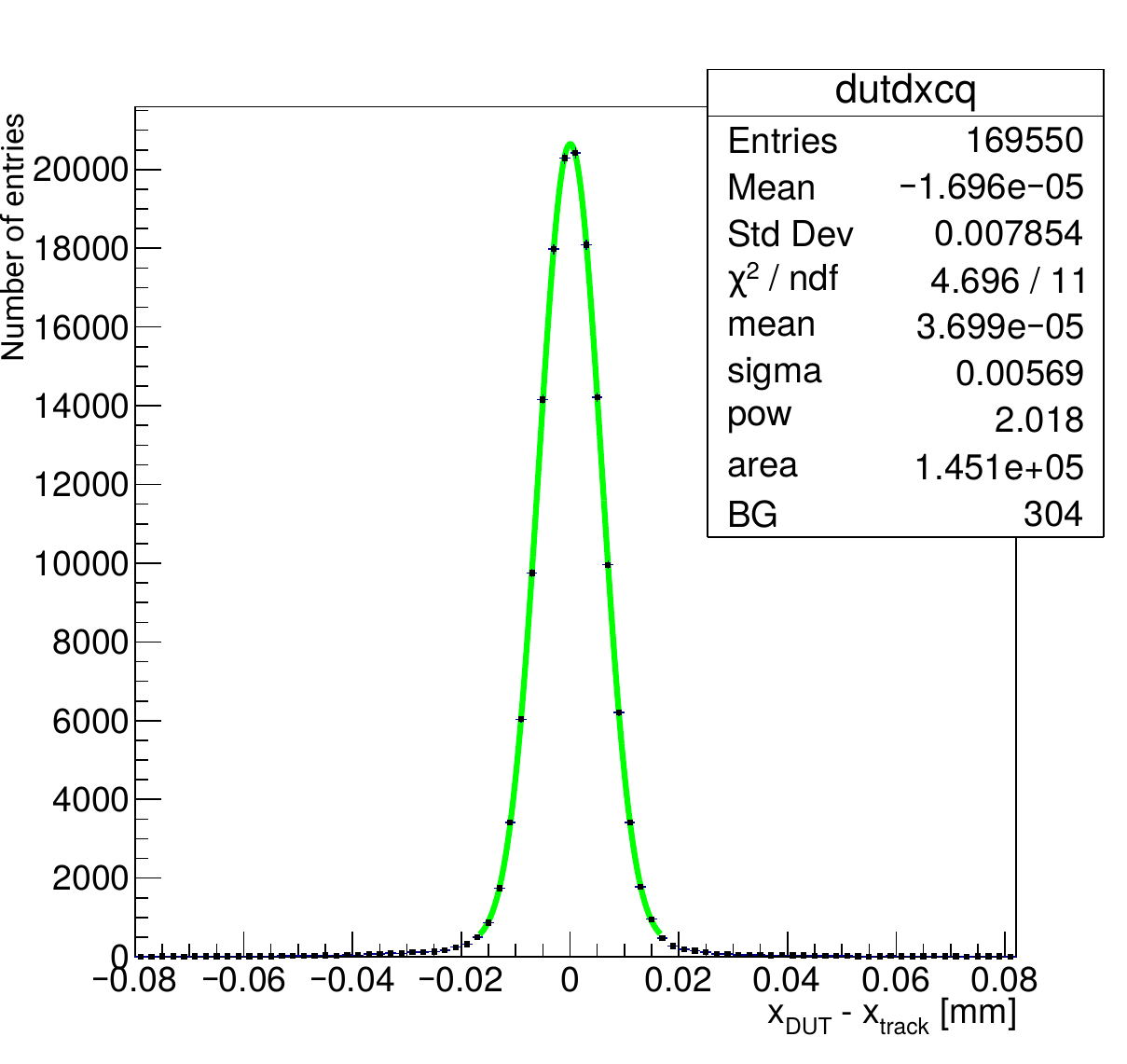}
\includegraphics[width=0.4\linewidth]{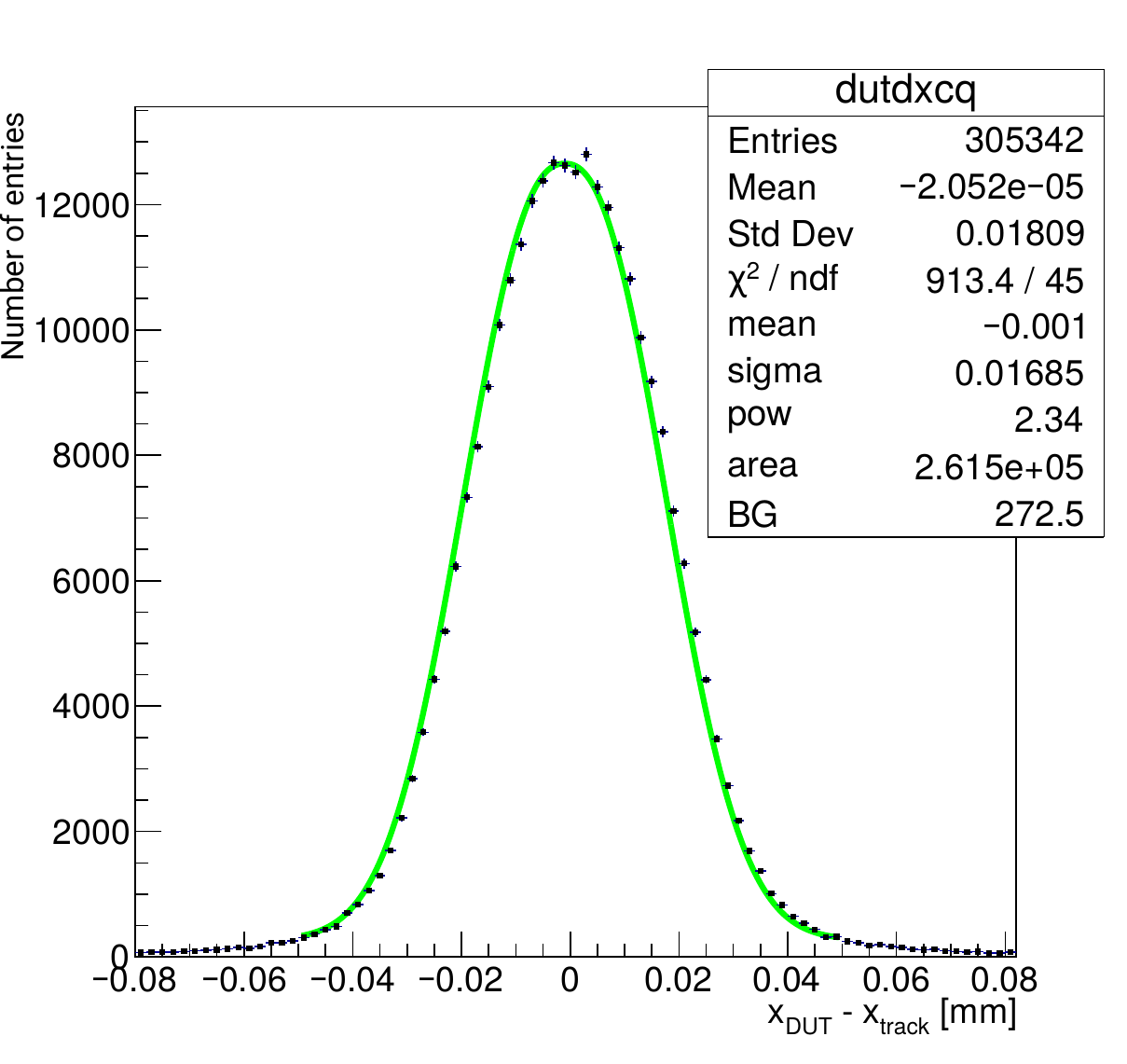}
\includegraphics[width=0.4\linewidth]{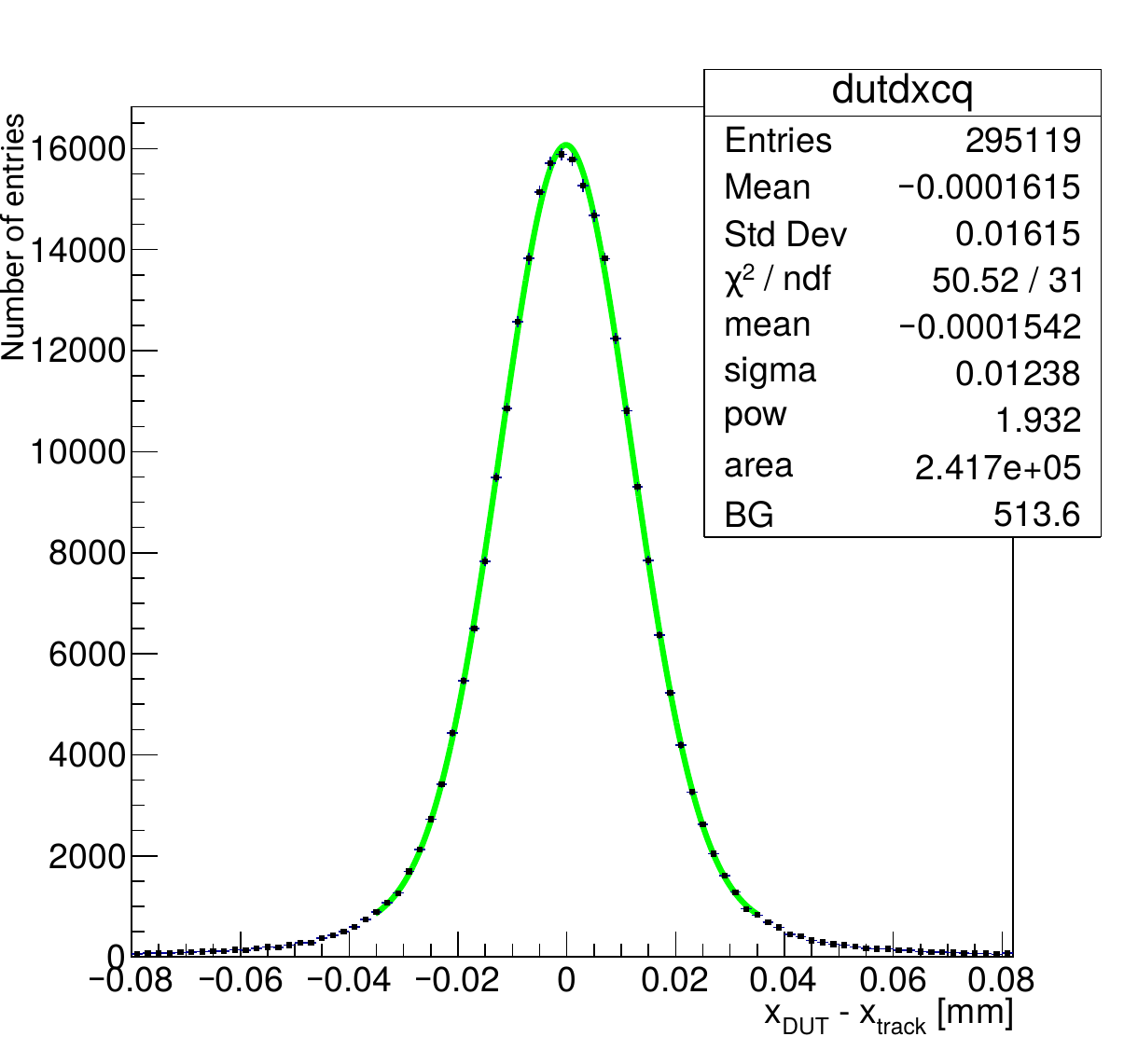}
    \caption{Residual distributions fitted with generalized error functions at vertical incidence (left) and at the optimal angle (right) for 
    modules 
    M563 (\SI{100x25}{\micro\meter} sensor, non-irradiated) (top),
    M564 (\SI{50x50}{\micro\meter} sensor, non-irradiated) (center),
    and M521 (\SI{50x50}{\micro\meter} sensor, $\Phi_{\rm eq} = $ \SI{5.3e15}{\per\centi\meter\squared}) (bottom).
    The fit parameters listed under $\chi^{2}$ / ndf correspong to the parameres $\mu_{\rm g}$, $\sigma_{\rm g}$, $S_{\rm g}$, $A_{\rm g}$ and $B_{\rm g}$ in equation \ref{generalizedGaussian}, respectively.}
\label{fig_resplots}
\end{figure}

To determine the spatial resolution of a pixel module, first the distance in the local $x$ coordinate of a cluster in the DUT ($x_{\text{DUT}}$) to the impact point of the telescope track extrapolated to the DUT ($x_{\text{track}}$) is calculated for each event as:
\begin{equation}
\centering
    \Delta x = x_{\text{DUT}} - x_{\text{track}.}
\end{equation}
Then the resulting residual distribution is fitted using a generalized Gaussian distribution~\cite{generror} in a range of three standard deviations from the peak of the distribution:
\begin{equation}
    \label{generalizedGaussian}
\centering
    g_{\text {gen}}(\Delta y)  = B_{g} + \frac{A_{g}\cdot S_{g}}{\sqrt{8}\cdot\sigma_{g}\cdot\Gamma(1/S_{g})}\cdot\text{exp}\left(-\left| \frac{\Delta y - \mu_{g}}{\sqrt{2}\cdot\sigma_{g}}\right|^{S_{g}}\right)
\end{equation}
where $\Gamma$ is the Gamma function and all variables with a $g$ subscript are fit parameters.

More recent studies have used the RMS of the residual distribution 
restricted to a certain range instead of a fit to evaluate the spatial resolution. This method is preferred for angles close to vertical incidence where the distributions are box-like and not Gaussian. 
However, using the RMS requires additional cuts not used in this analysis to reject outliers, which would strongly bias the RMS.
The spatial resolution of the pixel module in $x$ direction is obtained 
from the standard deviation of the fitted generalized error function by subtracting the telescope 
resolution in quadrature.

The spatial resolution as a function of the angle of incidence for \SI{50x50}{\micro\meter} pitch sensors before and after irradiation and for a non-irradiated \SI{100x25}{\micro\meter} pitch sensor is shown in Fig.~\ref{fig_compRes}, while selected residual distributions fitted with generalized error functions are shown in Fig.~\ref{fig_resplots}.
The non-irradiated modules were measured at room temperature at a bias voltage of \SI{120}{\volt}. The irradiated module was measured at a temperature below \SI{-25}{\celsius} and a bias voltage of \SI{800}{\volt}. The results shown in Fig.~\ref{fig_compRes} are after subtraction of the telescope resolution in quadrature. 
The telescope resolution depends on the beam momentum, the spacing of the telescope planes, and the distance between the DUT and the neighboring telescope planes. In case of the non-irradiated modules, the telescope resolution is estimated 
using upstream and downstream triplet tracks, extrapolated to the DUT plane.
It is \SI[separate-uncertainty = true,multi-part-units = single]{4.3 \pm 0.5}{\micro\meter} and \SI[separate-uncertainty = true,multi-part-units = single]{3.8 \pm 0.5}{\micro\meter} for the setup with the module with \SI{50x50}{\micro\meter} and \SI{100x25}{\micro\meter} pixel size, respectively. 
This small difference is simply a result of two independent measurements with slightly different positions of the DUT with respect to the closest telescope planes.
For the irradiated module a telescope resolution of \SI{7.6}{\micro\meter} is subtracted, using only the three upstream telescope planes. In this case, the telescope resolution is obtained by placing a non-irradiated module with known resolution in the cold box with the same telescope configuration and measuring the residual distribution.  
A systematic uncertainty of \SI{0.5}{\micro\meter} on the final results is estimated.


For the sensor modules with \SI{50x50}{\micro\meter} pixel size, a minimum hit resolution of \SI{3.4}{\micro\meter} (\SI{9.4}{\micro\meter}) is obtained at the optimal angle of incidence,
\begin{equation}
\centering
   \alpha_{\text {opt}}  = \arctan\left( \frac{\text{pitch}}{\text{thickness} }\right),
\end{equation}
before (after) irradiation. 
The optimal angle is the angle at which a track traverses exactly two pixels, which leads to the best resolution because of charge sharing.
The optimal angle before irradiation is approximately \SI{18}{\degree} and increases to about \SI{20}{\degree} after irradiation, indicating a reduction in the charge collection depth. At vertical incidence, the resolution is about \SI{14}{\micro\meter} in both cases, which is in agreement with the expected resolution of $\text{pitch}/\sqrt{12}$ in the absence of charge sharing within the uncertainties. 
%
%
For the non-irradiated sensor with \SI{100x25}{\micro\meter} pixel size, 
a minimum spatial resolution in the \SI{25}{\micro\meter} direction of \SI{2}{\micro\meter} is obtained at the optimal angle of \SI{8.5}{\degree}.

\section{Summary and Outlook}
Silicon pixel modules with planar sensors from Hamamatsu Photonics K.K. and RD53A readout chips have been built and evaluated with a particle beam. After threshold tuning, a minimum threshold of \SI{700}{e} for non-irradiated RD53A-sensor assemblies and approximately \SI{1200}{e} after irradiation to a fluence of up to $\Phi_{\rm eq} = $  \SI{2.1e16}{\per\centi\meter\squared} is obtained.
This fluence corresponds to the highest lifetime fluence for planar sensors in the IT, which would be reached in parts of ring~1 of the forward discs, assuming ring~1 is not replaced during the HL-LHC. 
The current sensor design results in a crosstalk of approximately 11\% in the \SI{100x25}{\micro\meter} pitch modules. Based on this result and on TCAD studies, a  modification of the design has been implemented in the next submission, which reduces the crosstalk by 30\%. 
A spatial resolution of \SI{2}{\micro\meter} is obtained for a non-irradiated  \SI{100x25}{\micro\meter} pitch sensor at the optimal angle of incidence. 
For a sensor module with \SI{50x50}{\micro\meter} pixel size, a spatial resolution of \SI{3.4}{\micro\meter} before irradiation and  \SI{9.4}{\micro\meter} 
at a fluence of $\Phi_{\rm eq} = $  \SI{5.3e15}{\per\centi\meter\squared} 
is reached.
An efficiency in excess of 99\% is achieved for modules irradiated to fluences up to $\Phi_{\rm eq} = $ \SI{2.1e16}{\per\centi\meter\squared}.
Assuming NIEL scaling and sufficient cooling performance, 
these results show that planar silicon sensor modules
would survive the entire 10 year HL-LHC running program 
corresponding to an integrated luminosity of  \SI{4000}{\per\femto\barn}.
No replacements are needed except for barrel layer~1, 
where 3D sensors will be used and for which at least one replacement is foreseen. A replacement of ring~1 of TFPX might be necessary if the cooling performance becomes critical or if too many readout channels become noisy.

The sensors without the punch-through bias dot evaluated in this paper fulfill the 
requirements for the CMS Inner Tracker upgrade in terms of hit efficiency. 
Guided by tracking and physics simulation of the entire tracking system, CMS has decided to use \SI{100x25}{\micro\meter} pitch sensors throughout the entire IT.

Recently, irradiations of CMS pixel prototype modules have been performed with \SI{23}{GeV} protons at the CERN Proton Synchrotron, resulting in a much reduced TID in the readout chip. Single chip assemblies with the final prototype of the CMS readout chip, CROCv1, have been tested before and after irradiation. 
The validation of full prototype modules consisting of flexible printed circuit boards and pixel sensors interconnected to two and four RD53A and CROCv1 readout chips, respectively, 
is  under way. These sensors feature larger pixels in the regions between readout chips to avoid dead regions. 
The prototyping program with CROCv1 single chip assemblies and full modules will be reported in future publications while CMS is preparing for the production of sensors, readout chips and modules for the CMS IT for Phase-2.

\section*{Acknowledgments}
The measurements leading to these results have been performed at the Test Beam Facility at DESY Hamburg (Germany), a member of the Helmholtz Association (HGF).
We also thank the teams at the irradiation facilities (the Karlsruhe Compact Cyclotron (KAZ) and Birmingham MC40) for their support.

The tracker groups gratefully acknowledge financial support from the following funding agencies: BMWFW and FWF (Austria); FNRS and FWO (Belgium); CERN; MSE and CSF (Croatia); Academy of Finland, MEC, and HIP (Finland); CEA and CNRS/IN2P3 (France); BMBF, DFG, and HGF (Germany); GSRT (Greece); NKFIA K124850, and Bolyai Fellowship of the Hungarian Academy of Sciences (Hungary); DAE and DST (India); INFN (Italy); PAEC (Pakistan); SEIDI, CPAN, PCTI and FEDER (Spain); Swiss Funding Agencies (Switzerland); MST (Taipei); STFC (United Kingdom); DOE and NSF (U.S.A.). This project has received funding from the European Union’s Horizon 2020 research and innovation program under the Marie Sk\l odowska-Curie grant agreement No 884104 (PSI-FELLOW-III-3i) and project AIDA-2020, GA no. 654168. Individuals have received support from HFRI (Greece).





\bibliography{rd53a}

\newcommand{\cmsAuthorMark}[1]
{\hbox{\textsuperscript{\normalfont#1}}}

\newpage

\section*{The Tracker Group of the CMS Collaboration}
\addcontentsline{toc}{section}{The Tracker Group of the CMS Collaboration}

{\setlength{\parindent}{0cm}
{\setlength{\parskip}{0.4\baselineskip}

\textcolor{black}{\textbf{Institut~f\"{u}r~Hochenergiephysik, Wien, Austria}\\*[0pt]
W.~Adam, T.~Bergauer, K.~Damanakis, M.~Dragicevic, R.~Fr\"{u}hwirth\cmsAuthorMark{1}, H.~Steininger}

\textcolor{black}{\textbf{Universiteit~Antwerpen, Antwerpen, Belgium}\\*[0pt]
W.~Beaumont, M.R.~Darwish\cmsAuthorMark{2}, T.~Janssen, H.~Rejeb~Sfar, P.~Van~Mechelen
% ...
}

\textcolor{black}{\textbf{Texas~A\&M~University, College~Station, USA}\\*[0pt]
R.~Eusebi}

\textcolor{black}{\textbf{Vanderbilt~University, Nashville, USA}\\*[0pt]
P.~D'Angelo, W.~Johns}\\

% ...
1: Also at Vienna University of Technology, Vienna, Austria \\
2: Also at Institute of Basic and Applied Sciences, Faculty of Engineering, Arab Academy for Science, Technology and Maritime Transport, Alexandria, Egypt \\
% ...


\end{document}